\makeatletter
\def\cl@chapter{}
\makeatother

\makeatletter
\newcommand\newtag[2]{#1\phantomsection\def\@currentlabel{#1}\label{#2}}
\makeatother

\RequirePackage{rotating}

 \documentclass[twocolumn,english]{svjour3}        
\usepackage[style = authoryear, maxbibnames=9, labelnumber, defernumbers = true, backend=bibtex]{biblatex}
\makeatletter
\if@twocolumn
  \renewcommand\normalsize{%
    \@setfontsize\normalsize\@xpt{12.5pt}%
    \abovedisplayskip=3 mm plus6pt minus 4pt
    \belowdisplayskip=3 mm plus6pt minus 4pt
    \abovedisplayshortskip=0.0 mm plus6pt
    \belowdisplayshortskip=2 mm plus4pt minus 4pt
    \let\@listi\@listI}%

  \renewcommand\small{%
    \@setfontsize\small{8.5pt}\@xpt
    \abovedisplayskip 8.5\p@ \@plus3\p@ \@minus4\p@
    \abovedisplayshortskip \z@ \@plus2\p@
    \belowdisplayshortskip 4\p@ \@plus2\p@ \@minus2\p@
    \def\@listi{\leftmargin\leftmargini
      \parsep 0\p@ \@plus1\p@ \@minus\p@
      \topsep 4\p@ \@plus2\p@ \@minus4\p@
      \itemsep0\p@}%
    \belowdisplayskip \abovedisplayskip}
\else
  \if@smallext
    \renewcommand\normalsize{%
      \@setfontsize\normalsize\@xpt\@xiipt
      \abovedisplayskip=3 mm plus6pt minus 4pt
      \belowdisplayskip=3 mm plus6pt minus 4pt
      \abovedisplayshortskip=0.0 mm plus6pt
      \belowdisplayshortskip=2 mm plus4pt minus 4pt
      \let\@listi\@listI}%

    \renewcommand\small{%
      \@setfontsize\small\@viiipt{9.5pt}%
      \abovedisplayskip 8.5\p@ \@plus3\p@ \@minus4\p@
      \abovedisplayshortskip \z@ \@plus2\p@
      \belowdisplayshortskip 4\p@ \@plus2\p@ \@minus2\p@
      \def\@listi{\leftmargin\leftmargini
        \parsep 0\p@ \@plus1\p@ \@minus\p@
        \topsep 4\p@ \@plus2\p@ \@minus4\p@
        \itemsep0\p@}%
      \belowdisplayskip \abovedisplayskip}
  \else
    \renewcommand\normalsize{%
      \@setfontsize\normalsize{9.5pt}{11.5pt}%
      \abovedisplayskip=3 mm plus6pt minus 4pt
      \belowdisplayskip=3 mm plus6pt minus 4pt
      \abovedisplayshortskip=0.0 mm plus6pt
      \belowdisplayshortskip=2 mm plus4pt minus 4pt
      \let\@listi\@listI}%

    \renewcommand\small{%
      \@setfontsize\small\@viiipt{9.25pt}%
      \abovedisplayskip 8.5\p@ \@plus3\p@ \@minus4\p@
      \abovedisplayshortskip \z@ \@plus2\p@
      \belowdisplayshortskip 4\p@ \@plus2\p@ \@minus2\p@
      \def\@listi{\leftmargin\leftmargini
        \parsep 0\p@ \@plus1\p@ \@minus\p@
        \topsep 4\p@ \@plus2\p@ \@minus4\p@
        \itemsep0\p@}%
      \belowdisplayskip \abovedisplayskip}
  \fi
\fi
\let\footnotesize\small
\makeatother

\usepackage{cmap}

\usepackage[ngerman,main=english]{babel}
\addto\extrasenglish{\languageshorthands{ngerman}\useshorthands{"}}

\usepackage{ifluatex}
\ifluatex
  \usepackage{fontspec}
  \usepackage[english]{selnolig}
\fi

\ifluatex
\else
  \usepackage[%
    rm={oldstyle=false,proportional=true},%
    sf={oldstyle=false,proportional=true},%
    tt={oldstyle=false,proportional=true,variable=false},%
    qt=false%
  ]{cfr-lm}
\fi

\ifluatex
\else
  \usepackage[T1]{fontenc}
  \usepackage[utf8]{inputenc} 
\fi

\smartqed  

\usepackage{graphicx}

\usepackage{upquote}

\usepackage{booktabs}

\usepackage{paralist}

\usepackage{csquotes}

\usepackage{textcmds}

\usepackage{tabularx}
\newcolumntype{Y}{>{\centering\arraybackslash}X}

\usepackage{xltabular}
\usepackage{multirow}
\usepackage{makecell}
\usepackage{caption}

\usepackage{url}

\usepackage{amsmath} 
\usepackage{bm}

\makeatletter
\g@addto@macro{\UrlBreaks}{\UrlOrds}
\makeatother

\usepackage[dvipsnames]{xcolor}

\usepackage[tikz]{ocgx2}
\usetikzlibrary{positioning,fit,arrows,shapes.geometric}

\usepackage{listings}
\renewcommand{\lstlistingname}{List.}
\usepackage{chngcntr}
\AtBeginDocument{\counterwithout{lstlisting}{section}}

\usepackage{stfloats}
\fnbelowfloat

\usepackage{hyperref}
\usepackage[all]{hypcap}

\usepackage[acronym,indexonlyfirst,nomain]{glossaries}
\glsdisablehyper

\usepackage[capitalise,nameinlink]{cleveref}
\crefname{section}{Sect.}{Sect.}
\Crefname{section}{Section}{Sections}
\crefname{listing}{\lstlistingname}{\lstlistingname}
\Crefname{listing}{Listing}{Listings}

\usepackage{xspace}

\DeclareFontFamily{U}{MnSymbolC}{}
\DeclareSymbolFont{MnSyC}{U}{MnSymbolC}{m}{n}
\DeclareFontShape{U}{MnSymbolC}{m}{n}{
  <-6>    MnSymbolC5
  <6-7>   MnSymbolC6
  <7-8>   MnSymbolC7
  <8-9>   MnSymbolC8
  <9-10>  MnSymbolC9
  <10-12> MnSymbolC10
  <12->   MnSymbolC12%
}{}
\DeclareMathSymbol{\powerset}{\mathord}{MnSyC}{180}

\usepackage{ifthen}
\usepackage{amssymb}
\usepackage[normalem]{ulem} 

\newboolean{showcomments}
\setboolean{showcomments}{false} 

\newboolean{showreview}
\setboolean{showreview}{false} 

\ifthenelse{\boolean{showcomments}}
{
	\usepackage{mparhack}

	\newcommand{\nbb}[3]{
		\marginpar[\hspace*{0.75cm}\parbox{35pt}{\tiny#1}]{\parbox{35pt}{\tiny#1}}
		{#2}
	}
	\newcommand{\modified}[1]{{\color{orange!80!black}#1}}
	\newcommand{\removed}[1]{{\color{red!90!black}\sout{#1}}}

	\newcommand{\rremoved}[2]{\nbb{#1}{\color{red!90!black} \sout{#2}}{red!90!black}}

}
{
	\newcommand{\nbb}[3]{}
	\newcommand{\modified}[1]{#1}
	
	\newcommand{\removed}[1]{}

	\newcommand{\rremoved}[2]{}

}

\usepackage{reviews-style}
\usepackage{pdfpages}

\hypersetup{hidelinks,
  colorlinks=true,
  allcolors=black,
  pdfstartview=Fit,
  breaklinks=true}

\journalname{JOURNALNAME}

\usepackage[math]{blindtext}
\usepackage{mwe}

\lstdefinelanguage{ATL}
{
	morekeywords={
		uses,
		if,
		module,
		for,
		create,
		from,
		refining,
		helper,
		context,
		def,
		rule,
		to,
		using,
		in,
		do,
		not,
		true,
		false,
		endif,
		else,
		then,
		Boolean,
		unique,
		lazy
	},
	sensitive=false, 
	morecomment=[l]{//}, 
	morecomment=[s]{/*}{*/}, 
	morestring=[b]" 
}
\lstdefinelanguage{QVTR}
{
	morekeywords={
		top,
		relation,
		domain
		String
	},
	sensitive=false, 
	morecomment=[l]{//}, 
	morecomment=[s]{/*}{*/}, 
	morestring=[b]" 
}

\usepackage{color}
\definecolor{eclipseBlue}{RGB}{42,0.0,255}
\definecolor{eclipseGreen}{RGB}{63,127,95}
\definecolor{eclipsePurple}{RGB}{127,0,85}

\definecolor{pblue}{rgb}{0.13,0.13,1}
\definecolor{pgreen}{rgb}{0,0.5,0}
\definecolor{pred}{rgb}{0.9,0,0}
\definecolor{pgrey}{rgb}{0.46,0.45,0.48}

\begin{document}

\title{Traceability and Reuse Mechanisms, the most important Properties of Model Transformation Languages 
}

\titlerunning{Traceability and Reuse Mechanisms, the most important Properties of Model Transformation Languages}        

%

\author{Stefan Höppner \and
        Matthias Tichy
}


\institute{S. Höppner \at
			  Ulm University\\
              James-Franck-Ring 1, 89081 Ulm \\
              Phone: +49-0731-5024208\\
              \email{stefan.hoeppner@uni-ulm.de}           
          \and
           M. Tichy \at
           Ulm University\\
              James-Franck-Ring 1, 80901 Ulm\\
              \email{matthias.tichy@uni-ulm.de}           
}

\date{Received: date / Accepted: date}

\maketitle

\begin{abstract}\hfill\break
\textbf{Context}\\
Dedicated model transformation languages are claimed to provide many benefits over the use of general purpose languages for developing model transformations.
However, the actual advantages and disadvantages associated with the use of model transformation languages are poorly understood empirically.
There is little knowledge and even less empirical assessment about what advantages and disadvantages hold in which cases and where they originate from.
In a prior interview study, we elicited expert opinions on what advantages result from what factors surrounding model transformation languages as well as a number of moderating factors that moderate the influence.

\noindent
\textbf{Objective}\\
We aim to quantitatively asses the interview results to confirm or reject the influences and moderation effects posed by different factors.
We further intend to gain insights into how valuable different factors are to the discussion so that future studies can draw on these data for designing targeted and relevant studies.

\noindent
\textbf{Method}\\
We gather data on the factors and quality attributes using an online survey.
To analyse the data and examine the hypothesised influences and moderations, we use universal structure modelling based on a structural equation model.
Universal structure modelling produces significance values and path coefficients for each hypothesised and modelled interdependence between factors and quality attributes that can be used to confirm or reject correlation and to weigh the strength of influence present.

\noindent
\textbf{Results}\\
We analyzed 113 responses.
The results show that the MTL capabilities Tracing and Reuse Mechanisms are most important overall.
Though the observed effects were generally 10 times lower than anticipated.
Additionally, we found that a more nuanced view of moderation effects is warranted.
Their moderating influence differed significantly between the different influences, with the strongest effects being 1000 times higher than the weakest.

\noindent
\textbf{Conclusion}\\
The empirical assessment of MTLs is a complex topic that cannot be solved by looking at a single stand-alone factor.
Our results provide clear indication that evaluation should consider transformations of different sizes and use-cases that go beyond mapping one elements attributes to another.
Language development on the other hand should focus on providing practical, transformation specific reuse mechanisms that allow MTLs to excel in areas such as maintainability and productivity compared to GPLs.

\end{abstract}

\section{Introduction}
\label{sec:intro}

Model driven engineering (MDE) envisions the use of model transformations as a main activity during development~\parencite{Sendall2003}.
When practising MDE, model transformations are used for a wide array of tasks such as manipulating and evolving models~\parencite{metzger2005systematic}, deriving artefacts like source code or documentation, simulating system behaviour or analysing system aspects~\parencite{Schmidt2006}.

Numerous dedicated model transformation languages (MTLs) of different form, aim and syntax~\parencite{kahani2019survey} have been developed to aid with model transformations.
Using MTLs is associated with many benefits compared to using general purpose languages (GPLs), though little evidence for this has been brought forth~\parencite{Goetz2020}.
The number of claimed benefits is enormous and includes, but is not limited to, better \textit{Comprehensibility}, \textit{Productivity} and \textit{Maintainability} as well as easier \textit{development} in general~\parencite{Goetz2020}.
The existence of such claims can partially be attributed to the advantages that are ascribed to domain specific languages (DSLs)~\parencite{Hermans2009,Johannes2009}.

In a prior systematic literature review, we have shown that it is still uncertain whether these advantages exist and where they arise from~\parencite{Goetz2020}.
Due to this uncertainty it is hard to convincingly argue the use of MTLs over GPLs for transformation development.
This problem is exacerbated when considering recent GPL advancements, like Java Streams, LINQ in C\# or advanced pattern matching syntax, that help reduce boilerplate code~\parencite{Hoeppner2021} and have put them back into the discussion for transformation development.
Even a community discussion held at the 12th edition for the International Conference on Model Transformations (ICMT'19) acknowledges GPLs as suitable contenders~\parencite{Burgueno2019}.
Moreover, the few existing empirical studies on this topic provide mixed and limited results.
Hebig et al. found no direct advantage for the development of transformations, but did find an advantage for the comprehensibility of transformation code in their limited setup~\parencite{Hebig2018}.
A study conducted by us, found that certain use cases favour the use of MTLs, while in others the versatility of GPLs prevails~\parencite{Hoeppner2021}.
Overall there exists a gap in knowledge in what the exact benefits of MTLs are, how strong their impact really is and what parts of the language they originate from.

To bridge this gap, we conducted an interview study with 56 experts from research and industry to discuss the topic of advantages and disadvantages of model transformation languages~\parencite{Hoeppner2022}.
Participants were queried about their views on the advantages and disadvantages of model transformation languages and the origins thereof.
The results point towards three main-areas that are relevant to the discussion, namely \textit{General Purpose Languages Capabilities}, \textit{Model Transformation Languages Capabilities} and \textit{Tooling}.
From the responses of the interviewees we identified which claimed MTL properties are influenced by which sub-areas and why.
They also provided us with insights on moderation effects on these interdependencies caused by different \textit{Use-Cases}, \textit{Skill \& Experience levels} of users and \textit{Choice of Transformation Language}.

All results of the interview study are qualitative and therefore limited in their informative value as they do not provide indication on the strength of influence between the involved variables.
It is also not clear whether the influence model is complete and whether the views pretended by the interview participants withstand community scrutiny.
Therefore they only represent an initial data set that requires a quantitative and detailed analysis.

In this paper, we report on the results of a study to \textit{confirm or deny} the interdependencies hypothesised from our interview results.
We provide \textit{quantification of the influence strengths} and \textit{moderation effects}.
To ensure a more complete theory of interactions, we also present the results of exploring interdependencies between factors and quality properties not hypothesised in the interviews.

Due to limited resources, this study focuses on the effects of \textit{MTL capabilities} (namely Bidirectionality, Incrementality, Mappings, Model Management, Model Navigation, Model Traversal, Pattern Matching, Reuse Mechanisms and Traceability) on \textit{MTL properties} (namely Comprehensibility, Ease of Writing, Expressiveness, Productivity, Maintainability and \modified{Reusability} and Tool Support) in the context of their \textit{uses-case} (namely bidirectional or unidirectional, incremental or non-incremental, meta-model sanity, meta-model, model and transformation size and semantic gap between input and output), the \textit{skills \& experience} of users and \textit{language choice}.
Further studies can follow the same approach and focus on different areas.
Descriptions for all MTL capabilities and MTL properties can be found in \Cref{sec:background} and thorough explanations can be found in our previous works \parencite{Goetz2020,Hoeppner2022}.

The goal of our study is to provide quantitative results on the influence strengths of interdependences between model transformation language \textit{Capabilities} and claimed \textit{Quality Properties} as perceived by users.
Additionally we provide data on the strength of moderation expressed by \textit{contextual properties}.
The study is structured around the hypothesised interdependencies between these variables, and their more detailed breakdown, extracted from our previous interview study.
Each presumed influence of a \textit{MTL capability} on a \textit{MTL property} forms one hypothesis which is to be examined in this study.
All hypotheses are extended with an assumption of moderation by the context variables.
The system of hypotheses that arises from these deliberations is visualised in a structure model, which forms the basis for our study.
The structure model is depicted in \Cref{fig:structure_model}.
The model shows exogenous variables on the left and right and endogenous variables at the centre.
Exogenous variables depicted in a ellipse with a dashed outline constitute the hypothesised moderating variables.

All hypotheses investigated in our study are of the form: \textit{``<MTL Property> is (positively or negatively) influenced by <MTL Capability>''}.
They are represented by arrows from exogenous variables on the left of \Cref{fig:structure_model} to endogenous variable at the centre.
A moderation on the hypothesised influence is assumed from all exogenous variables on the right of the figure connected to the considered endogenous variable.
In total we investigate 31 hypothesised influences, i.e. the number of outgoing arrows from the exogenous variables on the left of \Cref{fig:structure_model}.

\tikzstyle{factor} = [ellipse, draw, node distance=2cm, text width=6em, text centered, minimum height=5em]
\tikzstyle{mod-factor} = [ellipse, draw, node distance=2cm, text width=6em, text centered, minimum height=5em,dashed]
\tikzstyle{property} = [ellipse, rounded corners, draw, node distance=3.25cm, text width=6em, text centered, align=center, minimum height=5em,fill=white]

\tikzstyle{error} = [circle, draw, text centered, xshift=4em, yshift=1em]

\tikzstyle{l} = [draw, -latex']

\pgfdeclarelayer{background}
\pgfsetlayers{background,main}

\renewcommand{\labelitemi}{$\bullet$}

\begin{figure*}
	\begin{tikzpicture}
		\node[factor] (factor-bx) at (0,-2) {Bidirectionality $\xi_1$};
		\node[factor,below of=factor-bx] (factor-inc) {Incrementality $\xi_2$};
		\node[factor,below of=factor-inc] (factor-map) {Mappings $\xi_3$};
		\node[factor,below of=factor-map] (factor-man) {Model Management $\xi_4$};
		\node[factor,below of=factor-man] (factor-nav) {Model Navigation $\xi_5$};
		\node[factor,below of=factor-nav] (factor-trv) {Model Traversal $\xi_6$};
		\node[factor,below of=factor-trv] (factor-pm) {Pattern Matching $\xi_7$};
		\node[factor,below of=factor-pm] (factor-rm) {Reuse Mechanisms $\xi_8$};
		\node[factor,below of=factor-rm] (factor-trc) {Traceability $\xi_9$};
		
		\node[mod-factor] (mfactor-mms) at (15,0) {Meta-model Size $\xi_{13}$};
		\node[mod-factor,below of=mfactor-mms] (mfactor-ms) {Model Size $\xi_{14}$};
		\node[mod-factor,below of=mfactor-ms] (mfactor-ts) {Transformation Size$\xi_{15}$};
		\node[mod-factor,below of=mfactor-ts] (mfactor-bus) {Bidirectional Use$\xi_{18}$};
		\node[mod-factor,below of=mfactor-bus] (mfactor-mtl) {Language Choice $\xi_{10}$};
		\node[mod-factor,below of=mfactor-mtl] (mfactor-lsk) {Language Skills $\xi_{11}$};
		\node[mod-factor,below of=mfactor-lsk] (mfactor-exp) {Experience $\xi_{12}$};
		\node[mod-factor,below of=mfactor-exp] (mfactor-io) {I/O Semantic gap $\xi_{16}$};
		\node[mod-factor,below of=mfactor-io] (mfactor-msa) {Meta-model sanity $\xi_{17}$};
		\node[mod-factor,below of=mfactor-msa] (mfactor-ius) {Incremental Use $\xi_{19}$};
		
		\node[property] (prop-comp) at (7.5,-0.5) {Comprehensi-bility $\eta_1$};
		
		\node[error, above of=prop-comp] (comp-error) {$\zeta_1$};
		\path[l] (comp-error.south west) -- (prop-comp.45);
		
		\node[property,below of=prop-comp] (prop-eow) {Ease of Writing $\eta_2$};
		
		\node[error, above of=prop-eow] (eow-error) {$\zeta_2$};
		\path[l] (eow-error.south west) -- (prop-eow.45);
		
		\node[property,below of=prop-eow] (prop-ex) {Expressiveness $\eta_3$};
		
		\node[error, above of=prop-ex] (exp-error) {$\zeta_3$};
		\path[l] (exp-error.south west) -- (prop-ex.45);
		
		\node[property,below of=prop-ex] (prop-ts) {Tool Support $\eta_4$};
		
		\node[error, above of=prop-ts] (ts-error) {$\zeta_4$};
		\path[l] (ts-error.south west) -- (prop-ts.45);
		
		\node[property,below of=prop-ts] (prop-mtb) {Maintainability $\eta_5$};
		
		\node[error, above of=prop-mtb] (mtb-error) {$\zeta_5$};
		\path[l] (mtb-error.south west) -- (prop-mtb.45);
		
		\node[property,below of=prop-mtb] (prop-pro) {Productivity $\eta_6$};
		
		\node[error, above of=prop-pro] (pro-error) {$\zeta_6$};
		\path[l] (pro-error.south west) -- (prop-pro.45);
		
		\node[property,below of=prop-pro] (prop-reu) {Reusability $\eta_7$};
		
		\node[error, above of=prop-reu] (reu-error) {$\zeta_1$};
		\path[l] (reu-error.south west) -- (prop-reu.45);

		
		\path[l] (factor-bx.east) -- (prop-comp.west);
		\path[l] (factor-bx.east) -- (prop-eow.west);
		\path[l] (factor-bx.east) -- (prop-ex.west);
		\path[l] (factor-bx.east) -- (prop-mtb.west);
		\path[l] (factor-bx.east) -- (prop-pro.west);
		
		\path[l] (factor-inc.east) -- (prop-comp.west);
		\path[l] (factor-inc.east) -- (prop-eow.west);
		\path[l] (factor-inc.east) -- (prop-ex.west);
		
		\path[l] (factor-map.east) -- (prop-comp.west);
		\path[l] (factor-map.east) -- (prop-eow.west);
		\path[l] (factor-map.east) -- (prop-ex.west);
		\path[l] (factor-map.east) -- (prop-mtb.west);
		\path[l] (factor-map.east) -- (prop-reu.west);
		
		\path[l] (factor-man.east) -- (prop-comp.west);
		\path[l] (factor-man.east) -- (prop-eow.west);
		\path[l] (factor-man.east) -- (prop-ex.west);
		
		\path[l] (factor-nav.east) -- (prop-comp.west);
		\path[l] (factor-nav.east) -- (prop-eow.west);
		\path[l] (factor-nav.east) -- (prop-ex.west);
		
		\path[l] (factor-trv.east) -- (prop-comp.west);
		\path[l] (factor-trv.east) -- (prop-eow.west);
		\path[l] (factor-trv.east) -- (prop-ex.west);
		\path[l] (factor-trv.east) -- (prop-pro.west);
		
		\path[l] (factor-pm.east) -- (prop-comp.west);
		\path[l] (factor-pm.east) -- (prop-ex.west);
		\path[l] (factor-pm.east) -- (prop-pro.west);
		
		\path[l] (factor-rm.east) -- (prop-reu.west);
		
		\path[l] (factor-trc.east) -- (prop-comp.west);
		\path[l] (factor-trc.east) -- (prop-eow.west);
		\path[l] (factor-trc.east) -- (prop-ex.west);
		\path[l] (factor-trc.east) -- (prop-pro.west);
		
		\path[l] (mfactor-mms.west) -- (prop-comp.east);
		\path[l] (mfactor-mms.west) -- (prop-eow.east);
		
		\path[l] (mfactor-ms.west) -- (prop-comp.east);
		\path[l] (mfactor-ms.west) -- (prop-eow.east);
		
		\path[l] (mfactor-ts.west) -- (prop-comp.east);
		\path[l] (mfactor-ts.west) -- (prop-eow.east);
		
		\path[l] (mfactor-bus.west) -- (prop-comp.east);
		\path[l] (mfactor-bus.west) -- (prop-eow.east);
		\path[l] (mfactor-bus.west) -- (prop-ex.east);
		\path[l] (mfactor-bus.west) -- (prop-mtb.east);
		\path[l] (mfactor-bus.west) -- (prop-pro.east);
		
		\path[l] (mfactor-mtl.west) -- (prop-comp.east);
		\path[l] (mfactor-mtl.west) -- (prop-eow.east);
		\path[l] (mfactor-mtl.west) -- (prop-ex.east);
		\path[l] (mfactor-mtl.west) -- (prop-ts.east);
		\path[l] (mfactor-mtl.west) -- (prop-mtb.east);
		\path[l] (mfactor-mtl.west) -- (prop-pro.east);
		\path[l] (mfactor-mtl.west) -- (prop-reu.east);
		
		\path[l] (mfactor-lsk.west) -- (prop-comp.east);
		\path[l] (mfactor-lsk.west) -- (prop-eow.east);
		\path[l] (mfactor-lsk.west) -- (prop-mtb.east);
		\path[l] (mfactor-lsk.west) -- (prop-pro.east);
		\path[l] (mfactor-lsk.west) -- (prop-reu.east);
		
		\path[l] (mfactor-exp.west) -- (prop-eow.east);
		\path[l] (mfactor-exp.west) -- (prop-mtb.east);
		
		\path[l] (mfactor-io.west) -- (prop-comp.east);
		\path[l] (mfactor-io.west) -- (prop-eow.east);
		\path[l] (mfactor-io.west) -- (prop-pro.east);
		
		\path[l] (mfactor-msa.west) -- (prop-eow.east);
		
		\path[l] (mfactor-ius.west) -- (prop-comp.east);
		\path[l] (mfactor-ius.west) -- (prop-eow.east);
		\path[l] (mfactor-ius.west) -- (prop-ex.east);
	\end{tikzpicture}
	\caption{Structure model depicting the hypothesised influence and moderation effects of factors on MTL properties.}
	\label{fig:structure_model}
\end{figure*}
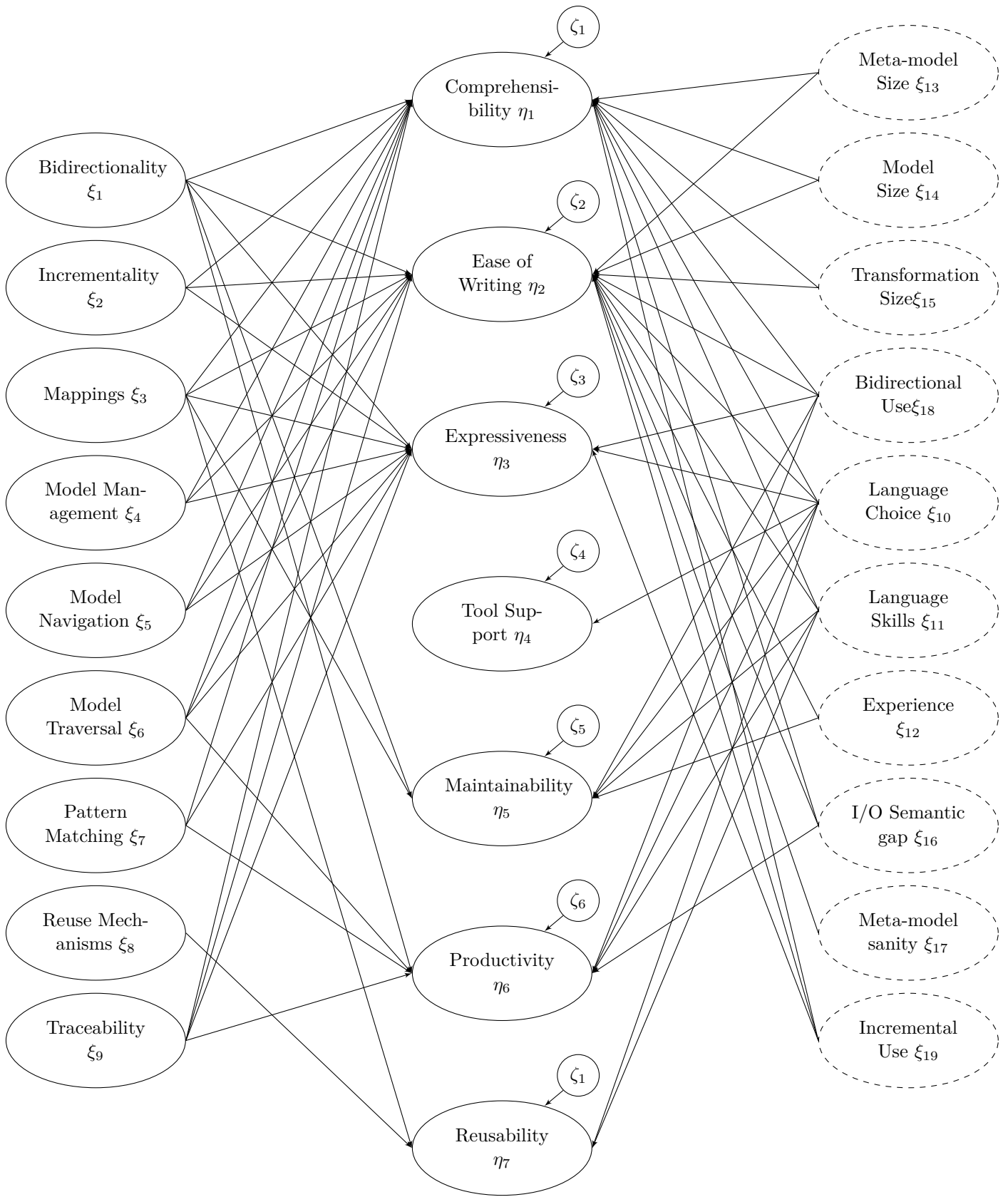

Our study is guided by the following research questions:

\begin{itemize}
	\item[\textbf{RQ1}] Which of the hypothesised interdependencies withstands a test of significance?
	\item[\textbf{RQ2}] How strong are the influences of model transformation language capabilities on the properties thereof?
	\item[\textbf{RQ3}] How strong are moderation effects expressed by the contextual factors \textit{use-case}, \textit{skills \& experience} and \textit{MTL choice}?
	\item[\textbf{RQ4}] What additional interdependencies arise from the analysis that were not initially hypothesised?
\end{itemize}

As the first study on this subject it contains confirmatory and exploratory elements.
We intend to confirm which of the interdependencies between \textit{MTL capabilities}, \textit{MTL properties} and \textit{contextual properties} withstand quantitative scrutiny (\textbf{RQ1}).
We explore how strong the influence and moderation effects between variables are (\textbf{RQ2 \& RQ3}), to gain new insights and to confirm their significance and relevance (minor influence strengths might suggest irrelevance even if goodness of fit tests confirm a correlation that is not purely accidental).
Lastly, we utilise the exploratory elements of USM to identify interdependencies not hypothesised by the experts in our interviews (\textbf{RQ4}).

We use an \textit{online survey} to gather data on language use and perceived quality of researchers and practitioners.
The responses are analysed using universal structure modelling (USM)~\parencite{buckler2008identifying} based on the structure model developed from the interview responses.
This results in a quantified structure model with influence weights, significance values and effect strengths.

Based on the responses from 113 participants, the key contributions of this paper are:

\begin{itemize}
	\item An adjusted structure model with newly discovered interdependencies;
	\item Quantitative data on the influence weight and effect strength of all factors as well as significant values for the influences;
	\item Quantitative data on the moderation strength of context factors;
	\item An analysis of the implications of the results for further empirical studies and language development;
	\item Reflections on the use of USM for investigating large hypotheses systems in software engineering research;
\end{itemize}

The method used in the reported study has been reviewed and published as part of the Registered Reports track at ESEM'22~\parencite{Hoeppner2022a}.

The structure of this paper is as follows:
\Cref{sec:background} provides an extensive overview of model-driven engineering, domain-specific languages, model transformation languages and structural equation modelling as well as universal structure modelling.
Afterwards, in \Cref{sec:methodology} the methodology is outlined.
Demographic data of the responses is reported in \Cref{sec:demographics} and the results of analysis is presented in \Cref{sec:results}.
In \Cref{sec:discussion} we discuss implications of the results and report our reflections on the use of USM.
\Cref{sec:threats} discusses threats to validity of our study and how we met them.
Lastly, in \Cref{sec:rw} we present related work before giving concluding remarks on our study in \Cref{sec:conclusion}.
\section{Background}
\label{sec:background}

In this section we provide the necessary background for our study.
Since it is a follow up study to our interview study~\parencite{Hoeppner2022} much of the background is the same and is therefore taken from those descriptions.
To stay self contained we still provide these descriptions.
This concerns \Cref{sec:background:mde,sec:background:dsl,sec:background:MTL}.
\Cref{sec:background:SEM,sec:background:descriptions} contains an extension of our descriptions from the registered report~\parencite{Hoeppner2022a}.

\subsection{Model-driven engineering}
\label{sec:background:mde}

The \textit{Model-Driven Architecture} (MDA) paradigm was first introduced by the Object Management Group in 2001~\parencite{OMG2001}.
It forms the basis for an approach commonly referred to as \textit{Model-driven development} (MDD) \parencite{Brown2005}, introduced as means to cope with the ever growing complexity associated with software development.
At the core of it lies the notion of using models as the central artefact for development.
In essence this means, that models are used both to describe and reason about the problem domain as well as to develop solutions~\parencite{Brown2005}.
An advantage ascribed to this approach that arises from the use of models in this way, is that they can be expressed with concepts closer to the related domain than when using regular programming languages~\parencite{Selic2003}.

When fully utilized, MDD envisions automatic generation of executable solutions specialized from abstract models~\parencite{Selic2003,Schmidt2006}.
To be able to achieve this, the structure of models needs to be known.
This is achieved through so called meta-models which define the structure of models.
The structure of meta-models themselves is then defined through meta-models of their own.
For this setup, the OMG developed a modelling standard called \textit{Meta-object Facility} (MOF)~\parencite{OMG2016} on the basis of which a number of modelling frameworks such as the \textit{Eclipse Modelling Framework} (EMF)~\parencite{steinberg2008emf} and the \textit{.NET Modelling Framework}~\parencite{hinkel2016nmf} have been developed.

\subsection{Domain-specific languages}
\label{sec:background:dsl}

Domain-specific languages (DSLs) are languages designed with a notation that is tailored for a specific domain by focusing on relevant features of the domain~\parencite{vanDeuersen2002}.
In doing so DSLs aim to provide domain specific language constructs, that let developers feel like working directly with domain concepts thus increasing speed and ease of development~\parencite{Sprinkle2009}.
Because of these potential advantages, a well defined DSL can provide a promising alternative to using general purpose tools for solving problems in a specific domain.
Examples of this include languages such as \textit{shell scripts} in Unix operating systems~\parencite{Kernighan1984}, \textit{HTML}~\parencite{Raggett1999} for designing web pages or AADL an architecture design language~\parencite{SAEMobilus2004}.

\subsection{Model transformation languages}
\label{sec:background:MTL}

The process of (automatically) transforming one model into another model of the same or different meta-model is called \textit{model transformation} (MT).
They are regarded as being at the heart of Model Driven Software Development~\parencite{Sendall2003,metzger2005systematic}, thus making the process of developing them an integral part of MDD.
Since the introduction of MDE at the beginning of the century, a plethora of domain specific languages for developing model transformations, so called model transformation languages (MTLs), have been developed~\parencite{Arendt2010,Balogh2006,Jouault2006,P86,10.1007/978-3-642-38883-5_7,P94,P23}.
Model transformation languages are DSLs designed to support developers in writing model transformations.
For this purpose, they provide explicit language constructs for tasks involved in model transformations such as model matching.
There are various features, such as directionality or rule organization~\parencite{Czarnecki2006}, by which model transformation languages can be distinguished.
For the purpose of this paper, we will only be explaining those features that are relevant to our study and discussion in \Cref{sec:background:MTL:exvsin,sec:background:MTL:rules,sec:background:MTL:direction,sec:background:MTL:increment,sec:background:MTL:tracing,sec:background:MTL:rac,sec:background:MTL:navigation}.
\Cref{tbl:MTL_features} provides an overview over the presented features.

Please refer to \textcite{Czarnecki2006,kahani2019survey,Mens2006} for complete classification.

\begin{table*}
	\caption{MTL feature overview}
	\label{tbl:MTL_features}
	\begin{tabularx}{\textwidth}{l|l|X}
		\toprule
		\textbf{Feature} & \textbf{Characteristic} & \textbf{Representative Language}\\
		\midrule
		\midrule
		\multirow{2}{7em}{Embeddedness} & Internal & FunnyQT (Clojure), RubyTL (Ruby), NMF Synchronizations (C\#)\\
		\cmidrule{2-3}
		& External & ATL, Henshin, QVT\\
		\midrule
		\multirow{2}{7em}{Rules} & Explicit Syntax Construct & ATL, Henshin, QVT\\
		\cmidrule{2-3}
		& Repurposed Syntax Construct & NMF Synchronizations (Classes), FunnyQT (Macros)\\
		\midrule
		\multirow{2}{7em}{Location Determination} & Automatic Traversal & ATL, QVT\\
		\cmidrule{2-3}
		& Pattern Matching & Henshin\\
		\midrule
		\multirow{2}{7em}{Directionality} & Unidirectional & ATL, QVT-O\\
		\cmidrule{2-3}
		& Bidirectional & QVT-R, NMF Synchronisations\\
		\midrule
		\multirow{2}{7em}{Incrementality} & Yes & NMF Synchronizations\\
		\cmidrule{2-3}
		& No & QVT-O\\
		\midrule
		\multirow{2}{7em}{Tracing} & Automatic & ATL, QVT\\
		\cmidrule{2-3}
		& Manual & NMF Synchronizations\\
		\midrule
		\multirow{2}{8em}{Dedicated Model Navigation Syntax} & Yes & ATL (OCL), QVT (OCL), Henshin (implicit in rules)\\
		\cmidrule{2-3}
		& No & NMF Synchronizations, FunnyQT, RubyTL\\
	\end{tabularx}
\end{table*}

\subsubsection{External and Internal transformation languages}
\label{sec:background:MTL:exvsin}

Domain specific languages, and MTLs by extension, can be distinguished on whether they are embedded into another language, the so called host language, or whether they are fully independent languages that come with their own compiler or virtual machine.

Languages embedded in a host language are called \textit{internal} languages.
Prominent representatives among model transformation languages are \textit{FunnyQT}~\parencite{10.1007/978-3-642-38883-5_7} a language embedded in Clojure, \textit{NMF Synchronizations} and the \textit{.NET transformation language}~\parencite{P23} embedded in C\#, and \textit{RubyTL}~\parencite{Cuadrado2006} embedded in Ruby.

Fully independent languages are called \textit{external} languages.
Examples of external model transformation languages include one of the most widely known languages such as the \textit{Atlas transformation language} (ATL)~\parencite{Jouault2006}, the graphical transformation language Henshin~\parencite{Arendt2010} as well as a complete model transformation framework called VIATRA~\parencite{Balogh2006}.

\subsubsection{Transformation Rules}
\label{sec:background:MTL:rules}

\textcite{Czarnecki2006} describe rules as being \textit{``understood as a broad term that describes the smallest units of [a] transformation [definition]''}.
Examples for transformation rules are the rules that make up transformation modules in ATL, but also functions, methods or procedures that implement a transformation from input elements to output elements.

The fundamental difference between model transformation languages and general-purpose languages that originates in this definition, lies in dedicated constructs that represent rules.
The difference between a transformation rule and any other function, method or procedure is not clear cut when looking at GPLs.
It can only be made based on the contents thereof.
An example of this can be seen in \Cref{lst:Java:example}, which contains exemplary Java methods.
Without detailed inspection of the two methods it is not apparent which method does some form of transformation and which does not.

In a MTL on the other hand transformation rules tend to be dedicated constructs within the language that allow a definition of a \textit{mapping} between input and output (elements).
The example rules written in the model transformation language ATL in \Cref{lst:ATL:example} make this apparent.
They define mappings between model elements of type \texttt{Member} and model elements of type \texttt{Male} as well as between \texttt{Member} and \texttt{Female} using \textit{rules}, a dedicated language construct for defining transformation mappings.
The transformation is a modified version of the well known Families2Persons transformation case~\parencite{fam2per}.

\begin{lstlisting}[float=t,language=Java,caption=Example Java methods,label=lst:Java:example]
public void methodExample(Member m) {
  System.out.println(m.getFirstName());
}
public void methodExample2(Member m) {
  Male target = new Male();
  target.setFullName(m.getFirstName() + " Smith");
  REGISTRY.register(target);
}
\end{lstlisting}

\begin{lstlisting}[float=t,language=ATL,caption=Example ATL rules,label=lst:ATL:example]
rule Member2Male {
  from
    s : Member (not s.isFemale())
  to
    t : Male (
    	fullName <- s.firstName + ' Smith'
    )
}
	
rule Member2Female {
  from
    s : Member (s.isFemale())
  to
    t : Female (
      fullName = s.firstName + ' Smith'
      partner = s.companion
    )
}
\end{lstlisting}

\subsubsection{Rule Application Control: Location Determination}
\label{sec:background:MTL:rac}

Location determination describes the strategy that is applied for determining the elements within a model onto which a transformation rule should be applied \parencite{Czarnecki2006}.
Most model transformation languages such as ATL, Henshin, VIATRA or QVT~\parencite{QvT2016}, rely on some form of \textit{automatic traversal} strategy to determine where to apply rules.

We differentiate two forms of location determination, based on the kind of matching that takes place during traversal.
There is the basic \textit{automatic traversal} in languages such as ATL or QVT, where single elements are matched to which transformation rules are applied.
The other form of location determination, used in languages like Henshin, is based on \textit{pattern matching}, meaning a model- or graph-\textit{pattern} is matched to which rules are applied.
This does allow developers to define sub-graphs consisting of several model elements and references between them which are then manipulated by a rule.

The \textit{automatic traversal} of ATL applied to the example from \Cref{lst:ATL:example} will result in the transformation engine automatically executing the \texttt{Member2Male} on all model elements of type \texttt{Member} where the function \texttt{isFemale()} returns \texttt{false} and the \texttt{Member2Female} on all other model elements of type \texttt{Member}.

\begin{figure}
	\includegraphics[width=\linewidth]{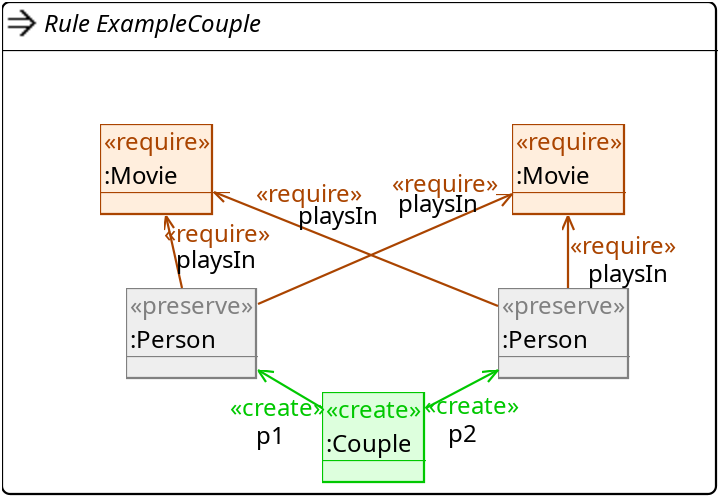}
	\caption{Example Henshin transformation}
	\label{fig:Henshin:example}
\end{figure}

The \textit{pattern matching} of Henshin can be demonstrated using \Cref{fig:Henshin:example}, a modified version of the transformation examples by \textcite{Krause2014}.
It describes a transformation that creates a couple connection between two actors that play in two films together.
When the transformation is executed the transformation engine will try and find instances of the defined graph pattern and apply the changes on the found matches.

This highlights the main difference between \textit{automatic traversal} and \textit{pattern matching} as the engine will search for a sub graph within the model instead of applying a rule to single elements within the model.

\subsubsection{Directionality}
\label{sec:background:MTL:direction}

The directionality of a model transformation describes whether it can be executed in one direction, called a unidirectional transformation or in multiple directions, called a multidirectional transformation~\parencite{Czarnecki2006}.

For the purpose of our study the distinction between unidirectional and bidirectional transformations is relevant.
Some languages allow dedicated support for executing a transformation both ways based on only one transformation definition, while other require users to define transformation rules for both directions.
General-purpose languages can not provide bidirectional support and also require both directions to be implemented explicitly.

The ATL transformation from \Cref{lst:ATL:example} defines a unidirectional transformation.
Input and output are defined and the transformation can only be executed in that direction.

The QVT-R relation defined in \Cref{lst:QVTR:example} is an example of a bidirectional transformation definition (For simplicity reasons the transformation omits the condition that males are only created from members that are not female).
Instead of a declaration of input and output, it defines how two elements from different domains relate to one another.
As a result given a \texttt{Member} element its corresponding \texttt{Male} elements can be inferred, and vice versa.

\begin{lstlisting}[float=t,language=QVTR,caption=Example QVT-R relation,label=lst:QVTR:example]
top relation Member2Male {
  n, fullName : String;
  domain Families s:Member {
    firstName = n };
  domain Persons t:Male {
    fullName = fullName};
  where {
  	fullName = n + ' Smith'; };
}
\end{lstlisting}

\subsubsection{Incrementality}
\label{sec:background:MTL:increment}

Incrementality of a transformation describes whether existing models can be updated based on changes in the source models without rerunning the complete transformation \parencite{Czarnecki2006}.
This feature is sometimes also called model synchronisation.

Providing incrementality for transformations requires active monitoring of input and/or output models as well as information which rules affect what parts of the models\modified{.
W}hen a change is detected the corresponding rules can then be executed.
It can also require additional management tasks to be executed to keep models valid and consistent.

\subsubsection{Tracing}
\label{sec:background:MTL:tracing}

According to \textcite{Czarnecki2006} tracing \textit{``is concerned with the mechanisms for recording different aspects of transformation execution, such as creating and maintaining trace links between source and target model elements''}.

Several model transformation languages, such as ATL and QVT have automated mechanisms for trace management.
This means that traces are automatically created during runtime.
Some of the trace information can be accessed through special syntax constructs while some of it is automatically resolved to provide seamless access to the target elements based on their sources.

An example of tracing in action can be seen in \texttt{line 16} of \Cref{lst:ATL:example}.
Here the \texttt{partner} attribute of a \texttt{Female} element that is being created, is assigned to \texttt{s.companion}.
The \texttt{s.companion} reference points towards a element of type \texttt{Member} within the input model.
When creating a \texttt{Female} or \texttt{Male} element from a \texttt{Member} element, the ATL engine will resolve this reference into the corresponding element, that was created from the referred \texttt{Member} element via either the \texttt{Member2Male} or \texttt{Member2Female} rule.
ATL achieves this by automatically tracing which target model elements are created from which source model elements.

\subsubsection{Dedicated Model Navigation Syntax}
\label{sec:background:MTL:navigation}

Languages or syntax constructs for navigating models is not part of any feature classification for model transformation languages.
However, it was often discussed in our interviews and thus requires an explanation as to what interviewees refer to.

Languages such as OCL~\parencite{OMG2014}, which is used in transformation languages like ATL, provide dedicated syntax for querying and navigating models.
As such they provide syntactical constructs that aid users in navigation tasks.
Different model transformation languages provide different syntax for this purpose.
The aim is to provide specific syntax so users do not have to manually implement queries using loops or other general purpose constructs.
OCL provides a functional approach for accumulating and querying data based on collections while Henshin uses graph patterns for expressing the relationship of sought-after model elements.

\subsection{Structural equation modelling and (Universal) Structural Equation Modelling}
\label{sec:background:SEM}

\begin{figure*}[ht]
	\includegraphics[scale=0.65]{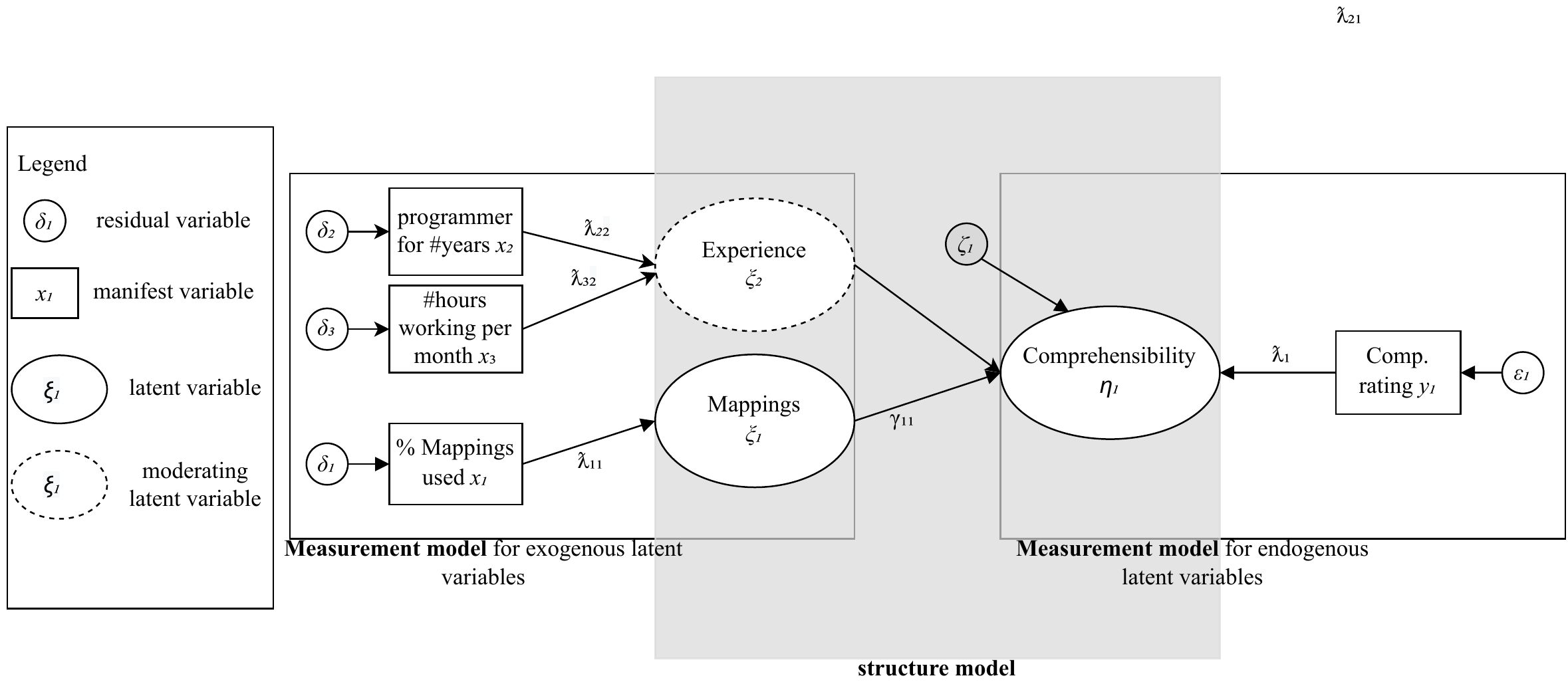}
	\caption{The makeup of a structural equation model.}~\label{fig:sample_SEM}
\end{figure*}

\textbf{Structural equation modelling (SEM)} is an approach used for confirmatory factor analysis~\parencite{10.1145/3469888}.
It defines a set of methods used to \textit{``investigate complex relationship structures between variables and allows for quantitative estimates of interdependencies thereof.
Its goal is to map the a-priori formulated cause-effect relationships into a linear system of equations and to estimate the model parameters in such a way that the initial data, collected for the variables, are reproduced as well as possible''}~\parencite{Weiber2021}.

Structural equation modelling distinguishes between two sets of variables \textit{manifest} and \textit{latent}.
\textit{Manifest} variables are variables that are empirically measured and \textit{latent} variables describe theoretical constructs that are hypothesised to interact with each other.
Latent variables are further divided into \textit{exogenous} or independent and endogenous or \textit{dependent} variables.

So called structural equation models, a sample of which can be seen in \Cref{fig:sample_SEM}, comprised of manifest and latent variables, form the heart of analysis.
They are made up of three connected sub-models.
The \textit{structure model}, the \textit{measurement model} of the exogenous latent variables and the \textit{measurement model} of the endogenous latent variables.

The \textit{structure model} defines all hypothesised interactions between exogenous ($\xi_{exID}$) and endogenous ($\eta_{endID}$) latent variables.
Each exogenous variable is linked, by arrow, to all endogenous variables that are presumed to be influenced by it.
Each of these connections is given a variable ($\gamma_{exID\_endID}$) that measures the influence strength.
If an exogenous variable \textit{moderates} the influences on a endogenous variable, the exogenous variable is depicted with a dashed outline and connected to all endogenous variables that are moderated by it\footnote{To illustrate moderation, arrows are usually shown from the moderating exogenous variable to the arrow representing the moderated influence , i.e., an arrow between an exogenous variable and an endogenous variable.
	However our illustration deviates from this due to the size and makeup of our hypothesis system.
	Standard representations can be found in the basic literature such as \textcite{Weiber2021}.}.
For each moderated influence a separate variable of the form $\gamma_{exID\_endID\_modEndID}$ is assigned.
In addition, an residual (or error) variable is appended to each endogenous latent variable to represent the influence of variables not represented in the model.

\Cref{fig:sample_SEM} shows an example structure equation model model for the hypothesis that \textit{``Mappings help with the comprehensibility of transformations, depending on the developers experience.''}.
The structure model seen at the centre of the figure, is comprised of the exogenous latent variable $\xi_1$ (\textit{Mappings}), the moderating exogenous variable $\xi_2$ (\textit{Experience}), the endogenous latent variable $\eta_1$ (\textit{Comprehensibility}), a presumed influence of \textit{Mappings} on \textit{Comprehensibility} via $\gamma_{11}$ and the error variable $\zeta_1$.
Lastly the model also contains a moderation of Experience on all influences of Comprehensibility.
As described earlier, this moderation effect is assigned the variable $\gamma_{11\_2}$.
The moderation variables are not depicted in our graphical representation of the structure model because of their high number and associated visibility issues.

The \textit{measurement model} of the exogenous latent variables reflects the relationships between all exogenous latent variables and their associated manifest variables.
Each manifest variable is linked, by arrow, to all exogenous latent variables that are measured through it.
Each of these connections is given a variable that measures the indication strength of the manifest variable for the latent variable.
Additionally, an error variable for each manifest variable is introduced that represents measurement errors.
In \Cref{fig:sample_SEM}, the measurement model for exogenous latent variables, seen at the left of the figure, is comprised of the exogenous latent variables $\xi_1$ (\textit{Mappings}) and $\xi_2$ (\textit{Experience}), the manifest variables $x_1$ (\textit{\% of code using Mappings}), $x_2$ (\textit{number of years a person has been a programmer}) and $x_3$ (\textit{number of hours per month spent developing transformations}) their measurement accuracy for Mapping usage $\lambda_{11}$ and their measurement accuracy for Experience $\lambda_{22}$ and and $\lambda_{32}$ and the associated measurement error $\delta_1$ and $\delta_2$ and $\delta_3$.

The \textit{measurement model} of the endogenous latent variables reflects the relationships between all endogenous latent variables and their associated manifest variables.
It is structured the same way as the \textit{measurement model} of the exogenous latent variables.
In \Cref{fig:sample_SEM}, it is shown on the right of the figure.

Given a structural equation model and measurements for manifest variables, the SEM approach calls for estimating the influence weights and latent variables within the models.
This is done in alternation for the measurement models and the structure model until a predefined quality criterion is reached.
Traditional methods (covariance-based structural equation modeling \& partial least squares) use different mathematical approaches such as maximum-likelihood estimation or least squares~\parencite{Weiber2021} to estimate influence weights.

\textbf{Universal Structure Modeling} (USM) is an exploratory approach that complements the traditional confirmatory SEM methods \parencite{buckler2008identifying}.
It combines the iterative methodology of partial least squares with a Bayesian neural network approach using multilayer perceptron architecture.
USM derives a starting value for latent variables in the model via principal component analysis and then applies the Bayesian neural network to discover an optimal system of linear, nonlinear and interactive paths between the variables.
This enables USM to identify complex relationships that may not be detected using traditional SEM approaches including hidden structures within the data and highlights unproposed model paths, nonlinear relations among model variables, and moderation effects.

The primary measures calculated in USM are the `Average Simulated Effect' (ASE), `Overall Explained Absolute Deviation' (OEAD), `interaction effect' (IE) and `parameter significance'.
ASE measures the average change in the endogenous variable resulting from a one-unit change in the exogenous variable across all simulations.
OEAD assesses the degree of fit between the observed and simulated values of the endogenous variable, capturing the overall explanatory power of the model.
IE evaluates the extent to which the effect of one exogenous variable on the endogenous variable depends on the level of another variable.
Parameter Significance determines whether the estimated coefficients for each exogenous variable in the model are statistically significant at a predetermined level of confidence which indicated if the exogenous variable has a meaningful impact on the endogenous variable and is calculated through a bootstrapping routine~\parencite{mooney1993bootstrapping}.
These metrics together provide a comprehensive assessment of the performance and explanatory power of a USM model.

USM is recommended for use in situations where traditional SEM approaches may not be sufficient to fully explore the relationships between variables.
Using USM instead of traditional structural equation modelling approaches is suggested for studies where there are still uncertainties about the completeness of the underlying hypotheses system and for exploring non-linearity in the influences~\parencite{Weiber2021,buckler2008identifying}.
Moreover its use of a neural network also reduces the requirements for the scale levels of data thus allowing the introduction of categorical variables in addition to metric variables~\parencite{Weiber2021}.

At present, the tool NEUSREL\footnote{\url{https://www.neusrel.com}} is the only tool available for conducting USM.

\subsection{MTL Quality Properties}
\label{sec:background:descriptions}

There exists a large body of quality properties that get associated with model transformation languages.
In literature many claims are made about advantages or disadvantages of MTLs in these different properties.
We categorised these properties in a previous work of ours~\parencite{Goetz2020}.
This study focuses on a subset of all the identified quality properties of MTLs which requires them to be properly explained.
In this section, we give a brief description of our definitions of each of the quality properties of MTLs relevant to the study.

\textit{Comprehensibility} describes the ease of understanding the purpose and functionality of a transformation based on reading code.

\textit{Ease of Writing} describes the ease at which a developer can produce a transformation for a specific purpose.

\textit{Expressiveness} describes the amount of useful dedicated transformation concepts in a language.

\textit{Productivity} describes the degree of effectiveness and efficiency with which transformations can be developed and used.

\textit{Maintainability} describes the degree of effectiveness and efficiency with which a transformation can be modified.

\textit{Reusability} describes the ease of reusing transformations or parts of transformations to create new transformations (with different purposes).

\textit{Tool Support} describes the amount of quality tools that exist to support developers in their efforts.
\section{Methodology}
\label{sec:methodology}

The methodology used in this study has been reviewed and published as part of the Registered Reports track at ESEM'22~\parencite{Hoeppner2022a}.
In the following, we provide a more detailed description and highlight all deviations from the reported method as well as justification for the changes.

The study itself is comprised of the following steps which were executed sequentially and are reported on in this section.

\begin{enumerate}
	\item Development of survey methodology.
	\item Submission to the Registered Reports track at EMSE'22.
	\item Methodology revision based on feedback.
	\item Development of online survey using an on premise version of the survey tool LimeSurvey\footnote{\url{https://www.limesurvey.org/}}.
	\item Survey review and pilot test by co-authors.
	\item Reworking survey based on pilot test.
	\item Opening online survey to public.
	\item Reaching out to potential survey subjects per mail and social media.
	\item Closing of online survey (9 weeks after opening).
	\item Data extraction.
	\item Data analysis using the USM tool NEUSREL.
\end{enumerate}

The steps executed differ in two ways from those reported in the registered report.
First, we do not contact potential participants for a second time after two weeks.
This was deemed unnecessary based on the number of participants at that point in time.
Moreover we did not want to bother those that participated already and had no way of knowing their identity.
Second, we kept the survey open 3 weeks longer than intended due to receiving several requests to do so.

\subsection{Survey Design}

In this section we detail the design of the used questionnaire and methodology used to develop and distribute it.

\subsubsection{Questionnaire}
\label{sec:vars}

The questions in the questionnaire are designed to query data for measuring the latent variables from the structure model in \Cref{fig:structure_model}.
The complete questionnaire can be found in \Cref{apdx:questions}.
In the following, we describe each latent variable and explain how we measure it through questions in the questionnaire.

There are 26 latent variables relevant to our study.
Variables $\xi_{1..19}$ describe exogenous variables and $\eta_{1..7}$ describe endogenous variables.
Each latent variable is measured through one or more manifest variables.
Extending the structure model from \Cref{fig:structure_model} with the manifest variables produces the complete structural equation model evaluated in this study.
Note that USM reduces the requirements for the scale levels of data thus allowing the use of categorical variables in addition to metric variables~\parencite{Weiber2021}.

All latent variables related to \textit{MTL capabilities} ($\xi_{1..9}$) are associated with a single manifest variable $x_{1..9}$, which measures how frequently the participants utilized the MTL capabilities in their transformations.
This measurement is represented as a ratio ranging from 0\% to 100\%.
The higher the value of $x_{1..9}$, the more frequently the participants used the MTL capabilities in their transformations.
Similarly, latent variables related to \textit{MTL properties} ($\eta_{1..7}$) are associated with a single manifest variable $y_{1..7}$ which measures the perceived quality of the property on a 5-point likert scale (e.g., very good, good, neither good nor bad, bad, very bad).

The use of single-item scales is a debated topic.
We justify their usage for the described latent variables on multiple grounds.
First, the latent variables are of high complexity due to the abstract concepts they represent.
Second, our study aims to produce first results that need to be investigated in more detail in follow up studies, more focused on single aspects of the model.
And third, due to the size of our structural equation model multi-item scales for all latent variables would increase the size of the survey, potentially putting off many subjects.
The validity of these deliberations for using single-item scales is supported by \textcite{fuchs2009using}.

The latent variable \textit{language choice} ($\xi_{10}$) is measured by means of querying participants to list their 5 most recently used transformation languages.
In our registered report we planned to also request participants to give an estimate on the percentage of their respective use \% ($x_{10})$.
This was discarded during pilot testing as it was seen as unnecessarily prolonging the questionnaire.
Pilot testers had difficulties providing accurate data and questioned whether this data was actually used in analysis.

\textit{Language skills} ($\xi_{11}$) is measured through $x_{11}$ and $x_{12}$ for which participants are asked to give the amount of years they have been using each language ($x_{11}$) and the amount of hours they use the language per month ($x_{12}$).

Similarly, \textit{experience} ($\xi_{12}$) is associated with the amount of years subjects have been involved in defining model transformations ($x_{13}$) and the amount of hours they spend on developing transformations each month ($x_{14}$).

\textit{Meta-model size} ($\xi_{13}$) and \textit{model size} ($\xi_{14}$) both require participants to state the range between which their (meta-) models vary ($x_{15}$, $x_{16}$).
This is measured by offering participants a number of ranges of (meta-) model objects.
For each range participants should give an estimate on how much percent of the (meta-) models they work fall within that size range.
For models the ranges are: \#objects $\leq 10$, $10 \le$ \#objects $\leq 100$, $100 \le$ \#objects $\leq 1000$, $1000 \le$ \#objects $\leq 10000$, $10000 \le$ \#objects $\leq 100000$, $100000 \le$ \#objects.
For meta-model the ranges are: \#objects $\leq 10$,  $10 \le$ \#objects $\leq 20$, $20 \le$ \#objects $\leq 50$, $50 \le$ \#objects $\leq 100$, $100 \le$ \#objects $\leq 1000$, $1000 \le$ \#objects.
Similarly, \textit{Transformation size} ($\xi_{15}$) is measured on a range of lines of code ($x_{17}$).
The options being: LOC $\leq 100$, $100 \le$ LOC $\leq 500$, $500 \le$ LOC $\leq 1000$, $1000 \le$ LOC $\leq 5000$, $5000 \le$ LOC $\leq 10000$, $10000 \le$ LOC.
Querying size data in this manner and the associated ranges have been successfully applied in a prior work the authors were involved in~\parencite{JOT:issue_2021_02/article5}.

To formulate the \textit{semantic gap between input and output} ($\xi_{16}$) we elicit the similarity of the structure ($x_{18}$) and data types ($x_{19}$) on a 5-point likert scale (very similar, similar, neither similar nor dissimilar, dissimilar, very dissimilar).
Participants are asked to give the percentage of all their meta-models that fall within each of the five assessments.

The \textit{meta-model sanity} ($\xi_{17}$) is measured through means of how well participants perceive their structure ($x_{20}$) and their documentation ($x_{21}$) to be on a 5-point scale (very well, well, neither well nor bad, bad, very bad).
Participants are asked to give the percentage of all their meta-models that fall within each of the five assessments.

Lastly, for both \textit{bidirectional uses} ($\xi_{18}$) and \textit{incremental uses} ($\xi_{19}$) we query participants on the ratio of bidirectional ($x_{22}$) and incremental ($x_{23}$) transformations compared to simple uni-directional transformations they have written.

\subsubsection{Pilot Study}
\label{sec:pilot}

We pilot tested the study with three researchers from the institute.
All pilot testers are researchers in the field of model driven engineering with more than 5 years of experience.
Based on their feedback, we reworded some questions questions, removed the usage percentage part of the question for \textit{language choice} and added more precise descriptions of the queried concepts.
We then made the questionnaire publicly available and distributed a link to it via emails.

\subsubsection{Target Subjects \& Distribution}
\label{sec:distribution}

The target subjects are both researchers and professionals from industry that have used dedicated model transformation languages to develop model transformations in the last five years.
We use voluntary and convenience sampling to select our study participants.
Both authors reached out to researchers and professionals they knew personally via mail and request them to fill out the online survey.
We further reach out, via mail, to all authors of publications listed in \textit{ACM Digital Library}, \textit{IEEE Xplore}, \textit{Springer Link} and \textit{Web of Science} that contain the key word \textit{model transformation} from the last five years.
A third source of subjects is drawn from social media.
The authors use their available social media channels to recruit further subjects by posting about the online-survey on the platforms.
The social media platform used for distribution was MDE-Net\footnote{\url{https://mde-network.com/}}, a community platform dedicated to model driven engineering.

The sampling method differs from the intended method by not including snowballing sampling as a secondary sampling method.
We decided on this to have more control over the subjects receiving a link to the study as we believe secondary and tertiary contacts might be too far secluded from our target subjects.

Participation was voluntary and we did not incentivise participation through offering rewards.
This decision is rooted in our experience in previous studies one other survey with 83 subjects \parencite{JOT:issue_2021_02/article5} and the interview study we are basing this study on with 56 subjects \parencite{Hoeppner2022}.

It is suggested in literature to have between 5 to 10 times as many participants as the largest number of parameters to be estimated in each structural equation (i.e., the largest number of incoming paths for a latent model variable)~\parencite{buckler2008identifying}.
Thus, the minimal number of subjects for our study to achieve stable results is 80.
To gain any meaningful results a sample size of 30 must not be undercut~\parencite{buckler2008identifying}.

In total we contacted 2383 potential participants and got 113\footnote{This constitutes a response rate of $~4.8\%$. We do however not know how many responses are a result of our social media posting.} responses exceeding the minimum requirement for stable results.

\subsection{Data Analysis}
\label{sec:analysis_plan}

We use USM to examine the hypotheses system modelled by the structure model shown in \Cref{fig:structure_model}.
USM is chosen over its structural equation modelling alternatives due to it being able to better handle uncertainty about the completeness of the hypothesis system under investigation, it having more capabilities to analyse moderation effects and the ability to investigate non linear correlations~\parencite{Weiber2021}.

USM requires a declaration of an initial likelihood of an interdependence between two variables.
This is used as a starting point for calculating influence weights but can change over the course of calculation.
For this, \textcite{buckler2008identifying} suggest to only assign a value of 0 to those relationships that are known to be wrong.
We use the results of our interview study~\parencite{Hoeppner2022}, shown in the structure model, to assign these values.
For each path that is present in the model, we assume a likelihood of 100\%.
To check for interdependencies that might have been missed by interview participants, we also use a likelihood of 100\% for all missing paths between $\xi_{1..19}$ and $\eta_{1..7}$.
Our plan was to use a likelihood of 50\% for these interdependencies but the tool available to us only allowed for either 100\% or 0\% to be put as input.

The tool NEUSREL is used on the extracted empirical data and the described additional input to estimate path weights and moderation weights within the extended structure model, i.e., the structure model where each exogenous latent variable is connected to all endogenous latent variables.
It also runs significance tests via a bootstrapping routine~\parencite{buckler2008identifying,mooney1993bootstrapping} and produces the significance value estimates for each influence.
The following procedures are then followed to answer the research questions from \Cref{sec:intro}.

\textbf{RQ1}.
We reject all hypothesised influences, i.e., those present in our structure model in \Cref{fig:structure_model}, that do not pass the statistical significance test.
The threshold we set for this is $0.01$.
Moreover, we discard hypothesised influences with minimal effects strengths that are several magnitudes lower than the median influence of all coefficients.
If, for example, the median of all path coefficients is 0.03 all influences with a coefficient lower or equal to 0.0009 are discarded.
We do so because such low influences suggest that the influence is negligible.

\textbf{RQ2 \& RQ3}.
All path coefficients produced that were not rejected in \textbf{RQ1} will then provide direct values for the influence and moderation strengths to answer \textbf{RQ2} \& \textbf{RQ3}.

\textbf{RQ4}.
The same significance criteria we applied to all hypothesised influences for \textbf{RQ1}, we also apply to the extended influences, i.e., those not present in the structure model from \Cref{fig:structure_model}.
Those influences that pass the significance test are added to the initial structural model as newly discovered influences.

\subsection{Privacy and Ethical concerns}
\label{sec:method:privacy}

All participants were informed of the data collection procedure, handling of the data and their rights, prior to filling out the questionnaire.
Participation was completely voluntary and not incentivised through rewards.

During selection of potential participants the following data was collected and processed.
\begin{itemize}
	\item First \& last name.
	\item E-Mail address.
\end{itemize}

The questionnaire did not collect any sensitive or identifiable data.

All data collected during the study was not shared with any person outside of the group of authors.

The complete information and consent form can be found in \Cref{apdx:consent_form}.

The study design was not presented to an ethical board.
The basis for this decision are the rules of the German Research Foundation (DFG) on when to use a ethical board in humanities and social sciences\footnote{\url{https://www.dfg.de/foerderung/faq/geistes_sozialwissenschaften/}}.
We refer to these guidelines because there are none specifically for software engineering research and humanities and social sciences are the closest related branch of science for our research.
\section{Demographics}
\label{sec:demographics}

We detail the background and experience of the participants in our study in the following sections.

\subsection{Experience in developing model transformations ($\xi_{12}$)}

\begin{figure}
	\includegraphics[width=\linewidth]{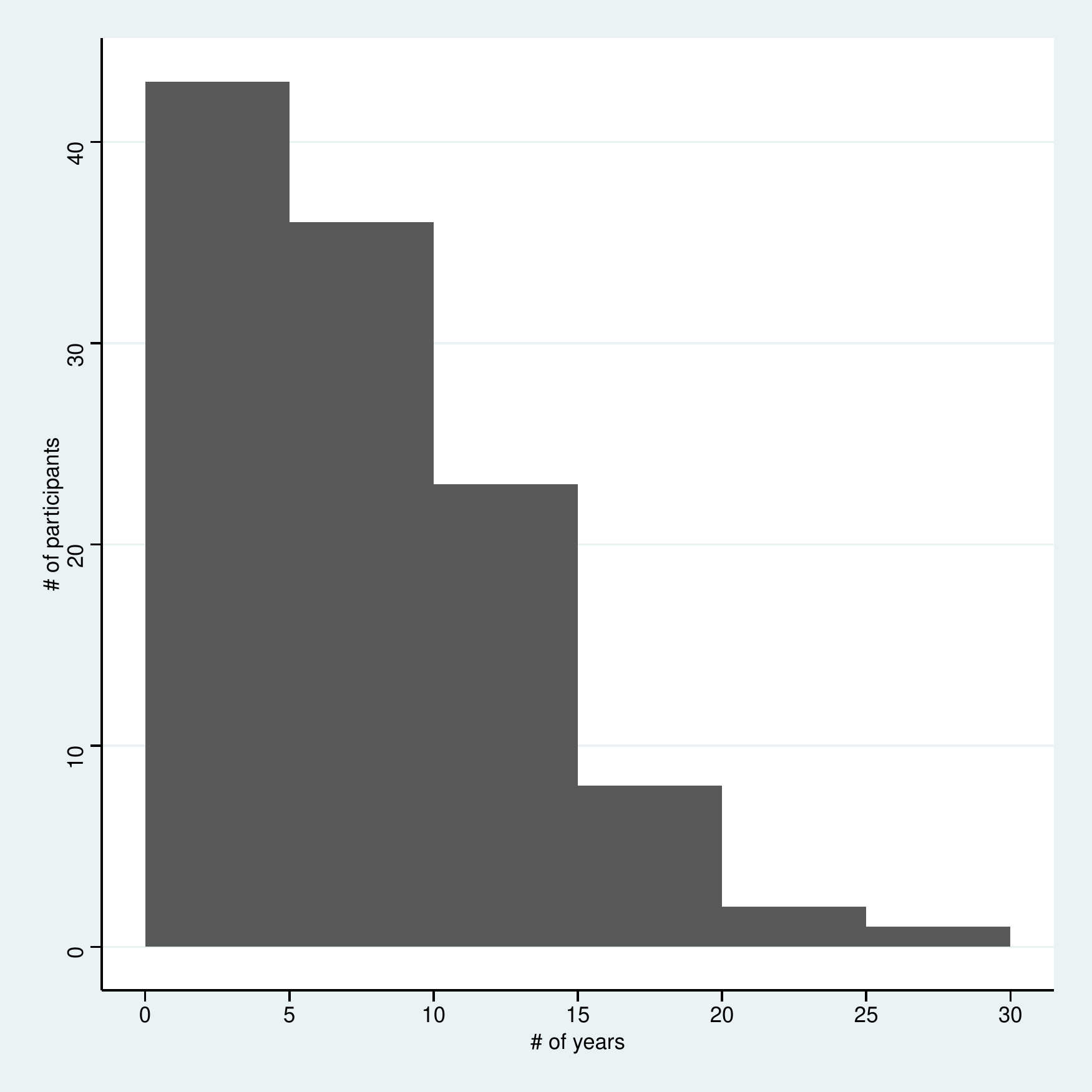}
	\caption{Histogram of participants total experience in years}\label{fig:exp_years}
\end{figure}

Our survey captured model transformation developers with wide range of experience.
The experience span ($x_{13}$) ranges from the least experience participant with half a year of experience up to the one with most experience of 30 years.
\Cref{fig:exp_years} shows a histogram of the experience stated by participants.
Over half of all participants have between 1 to ten years of experience in writing model transformations.
Three stated to have more than 20 years in total.
On average our participants have 9 years of experience.

\begin{figure}
	\includegraphics[width=\linewidth]{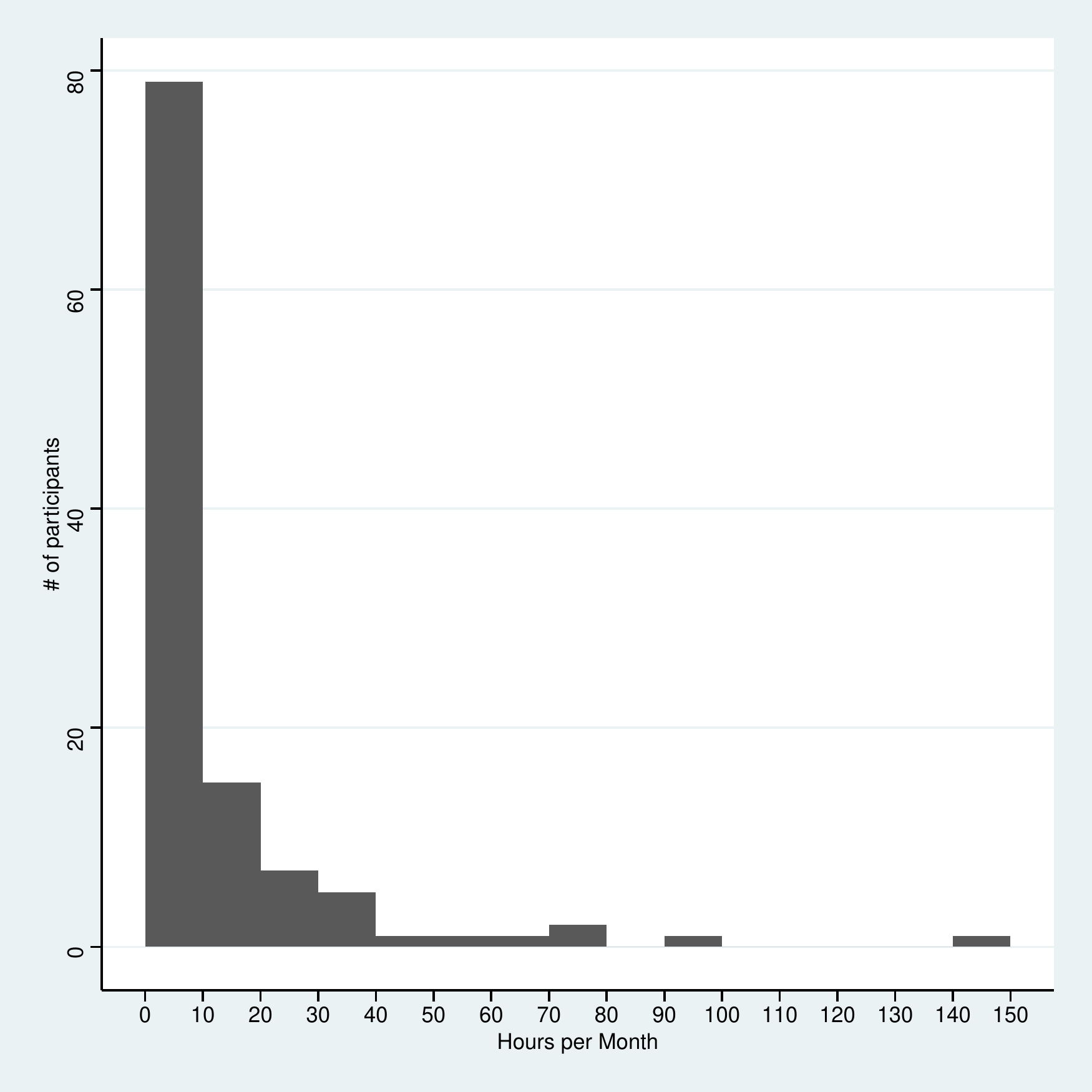}
	\caption{Histogram of participants recent experience in hours per month}\label{fig:exp_months}
\end{figure}

How much time participants spend developing transformations each month ($x_{14}$) also greatly varies.
Some participants have not developed transformations in recent time whereas others stated to spend 70 or more hours each month on transformation development.
\Cref{fig:exp_months} shows an overview over the hours participants spend each month in developing transformations.
The vast majority spends around 1 to 10 hours each month on transformation development.
Nine stated that they did not develop any transformation in recent times.
On average our participants spend about 14 hours per month developing model transformations.

\subsection{Languages used for developing model transformations ($\xi_{10}$) and experience therein ($\xi_{11}$)}

To develop their transformations, participants use a wide array of languages.
In total 43 languages ($x_{10}$) have been named 24 of which are unique languages used only by a single participant.

Surprisingly the language that has been used by the most participants is Java, a general purpose language.
Java has been used by 70 of the 113 participants.
The most used MTL is ATL with 58 users closely followed by another GPL, namely Xtend with 52 users.
\Cref{tbl:langs} shows how many participants use one of the ten most used languages for developing transformations.

\begin{table}
	\caption{Overview of languages used by participants}\label{tbl:langs}
	\centering
	\begin{tabular}{lr}
		\toprule
		Language & \# number of participants\\
		\midrule
		Java & 70\\
		ATL & 58\\
		Xtend & 52\\
		ETL & 29\\
		QVTo & 22\\
		Henshin & 14\\
		JavaScript & 12\\
		eMoflon & 7\\
		Fujaba & 5\\
		Python & 4\\
		\bottomrule
	\end{tabular}
\end{table}


Overall the prevalence of general purpose programming languages is higher than expected.
This might be explained by the large number of existing MTLs which reduce the amount of total users per language while only four different GPLs are used.

\subsection{Sizes ($\xi_{12}$,$\xi_{14}$)}

\begin{figure}
	\includegraphics[width=\linewidth]{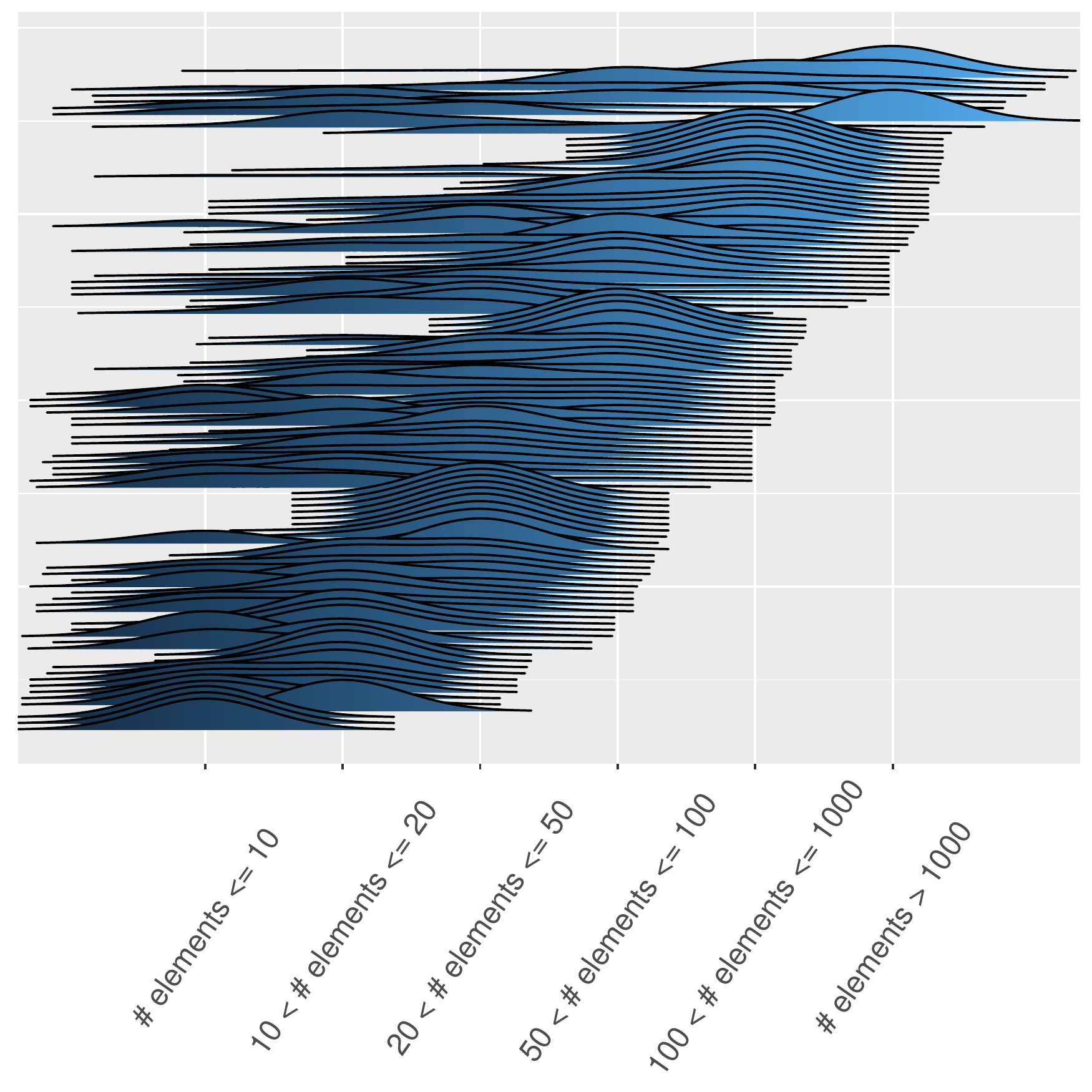}
	\caption{Distribution of meta-model sizes per participant}\label{fig:mm-sizes}
\end{figure}

The size distribution of meta-models ($x_{15}$) transformed by participants is shown in \Cref{fig:mm-sizes}.
On the x-axis the given intervals of meta-model sizes are shown and on the y-axis the distribution for each participant is shown.
For example, the first ridge line at the bottom of \Cref{fig:mm-sizes} shows the answers of a participant who has stated that 100\% of their transformations revolve around meta-models with 10 or less meta-model elements.

The figure illustrates that most transformations involve meta-models with 20 to 100 meta-model elements.
Moreover, most participants have some experience with small meta-models while only a handful of them has experience with transformations involving large meta models of more than 1.000 elements.

\begin{figure}
	\includegraphics[width=\linewidth]{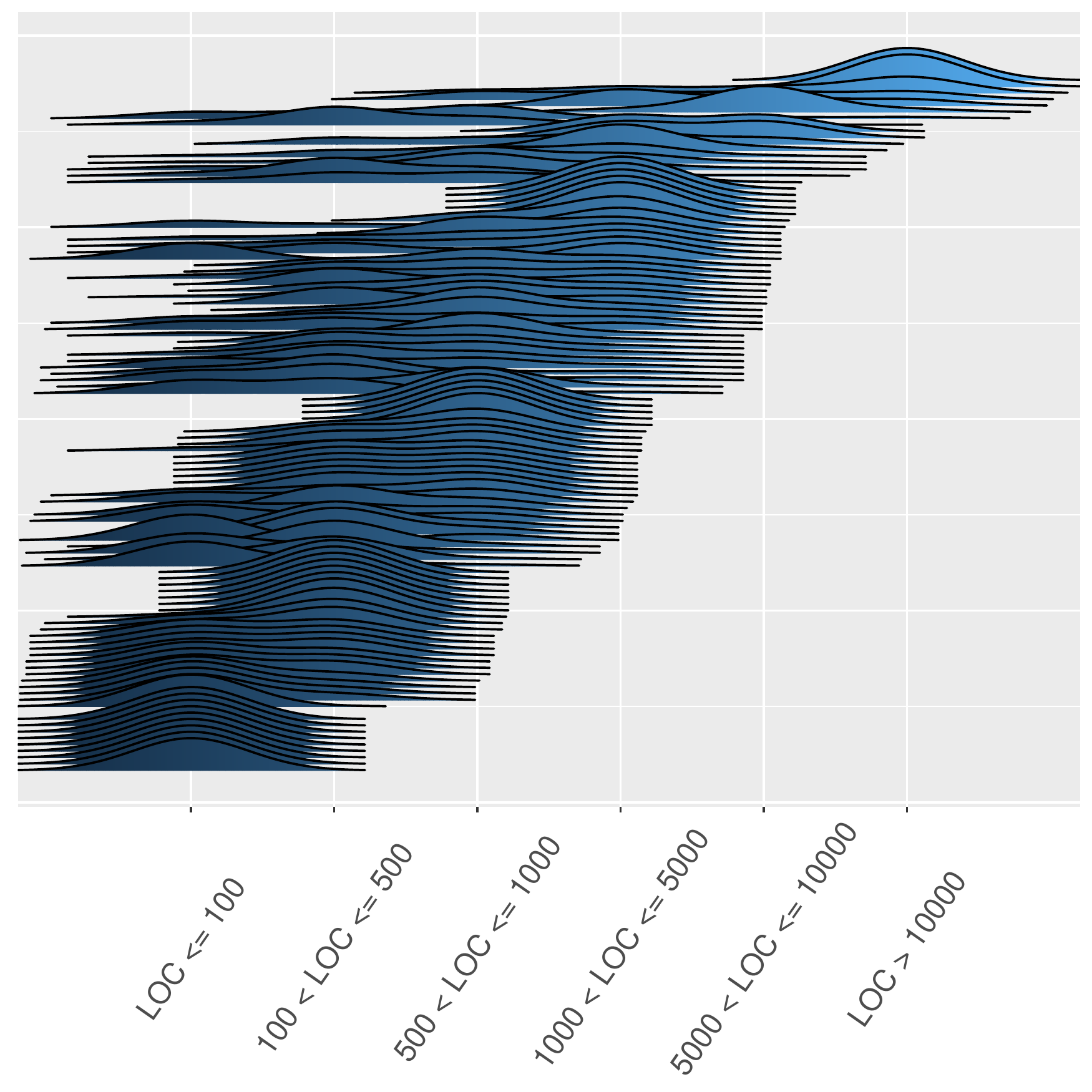}
	\caption{Distribution of transformation sizes per participant}\label{fig:t-sizes}
\end{figure}

The size distribution of model transformations ($x_{17}$) written by participants is shown in \Cref{fig:t-sizes}.
Similarly to the meta-model sizes, the figure illustrates that most participants have some experience with small transformations of sizes up to 100 lines of code.
Most also have experience with large transformations up to 1.000 lines of code.
More than 25\% of all participants also have experience with large and very large transformations ranging from 5.000 up to more than 10.000 lines of transformation code.

Overall the experience of our participants includes many moderately large to large transformations.
This strengthens us in the assumption that their answers are meaningful for our study.

\subsection{Conceptual distance between meta-models ($\xi_{16}$)}

\begin{figure}
	\includegraphics[width=\linewidth]{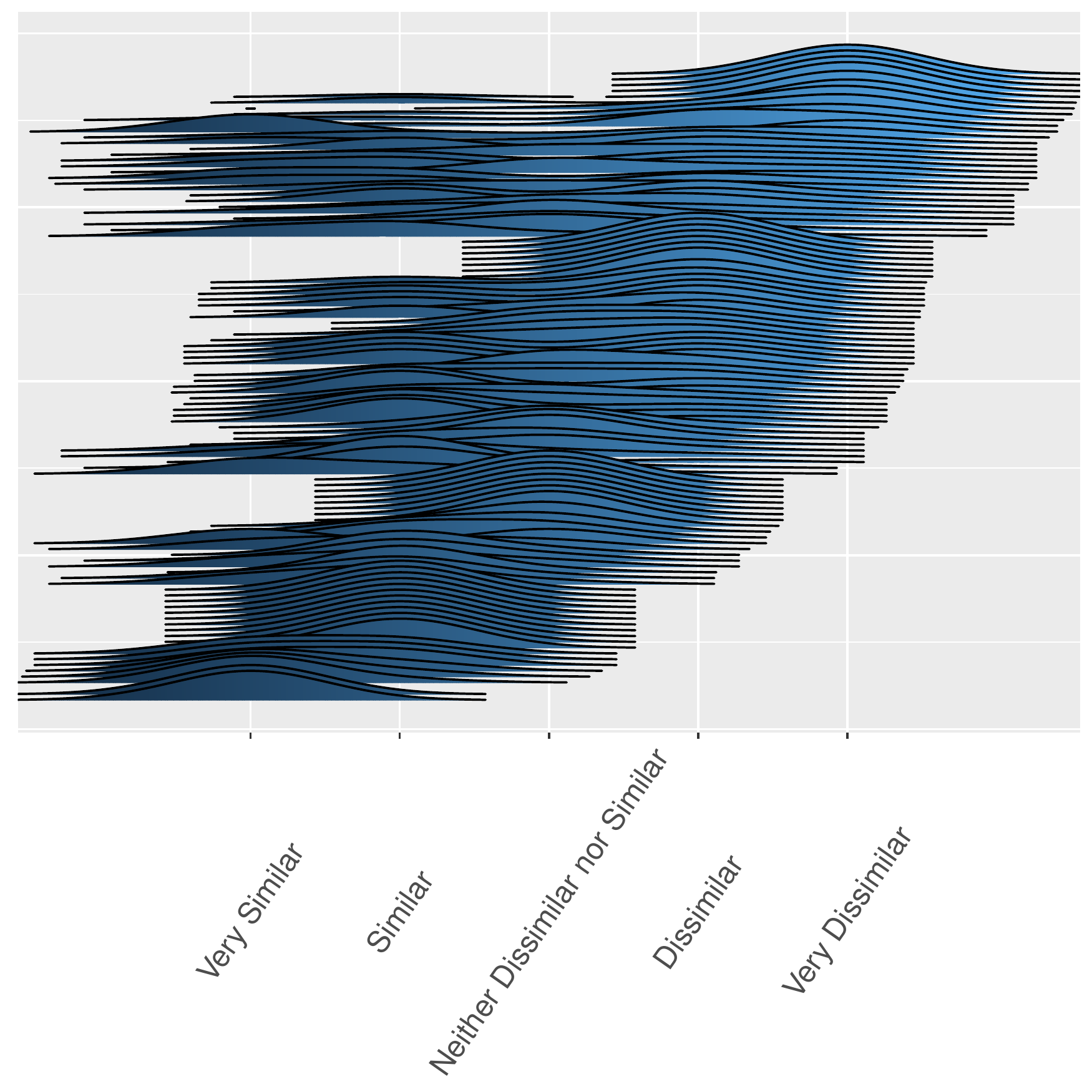}
	\caption{Distribution of input output meta-model structure similarity}\label{fig:mm-structure-sim}
\end{figure}

\begin{figure}
	\includegraphics[width=\linewidth]{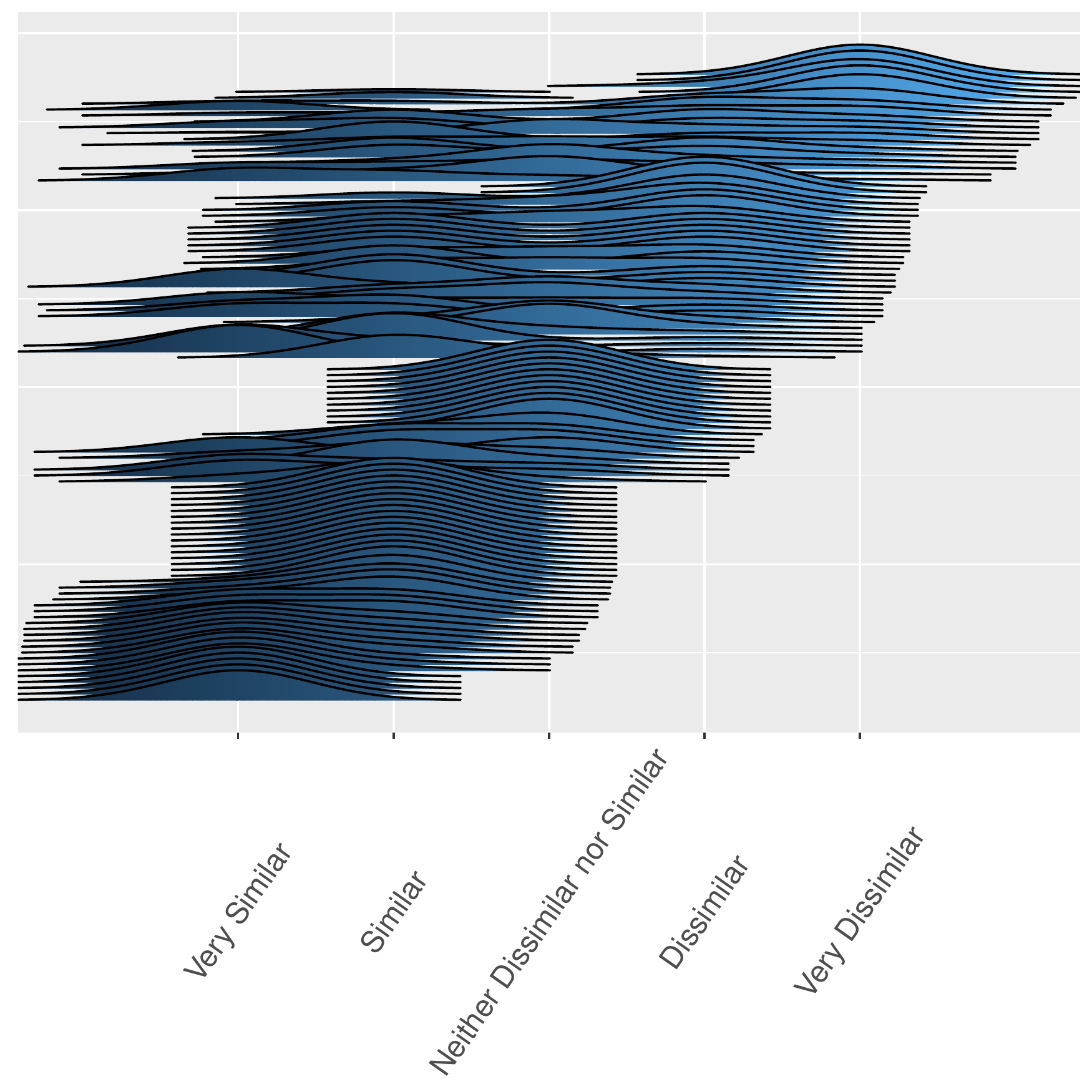}
	\caption{Distribution of input output meta-model attribute types similarity}\label{fig:mm-attr-sim}
\end{figure}

The similarity distribution of meta-models involved in the transformations of our participants is shown in \Cref{fig:mm-structure-sim} for the similarity of meta-model structures ($x_{18}$) and \Cref{fig:mm-attr-sim} for the similarity of data types ($x_{19}$).
Both show a even mix between structurally similar and distant meta-models as well as similar and dissimilar attribute types within the elements that are transformed into each other.

\subsection{Meta-model quality ($\xi_{17}$)}

\begin{figure}
	\includegraphics[width=\linewidth]{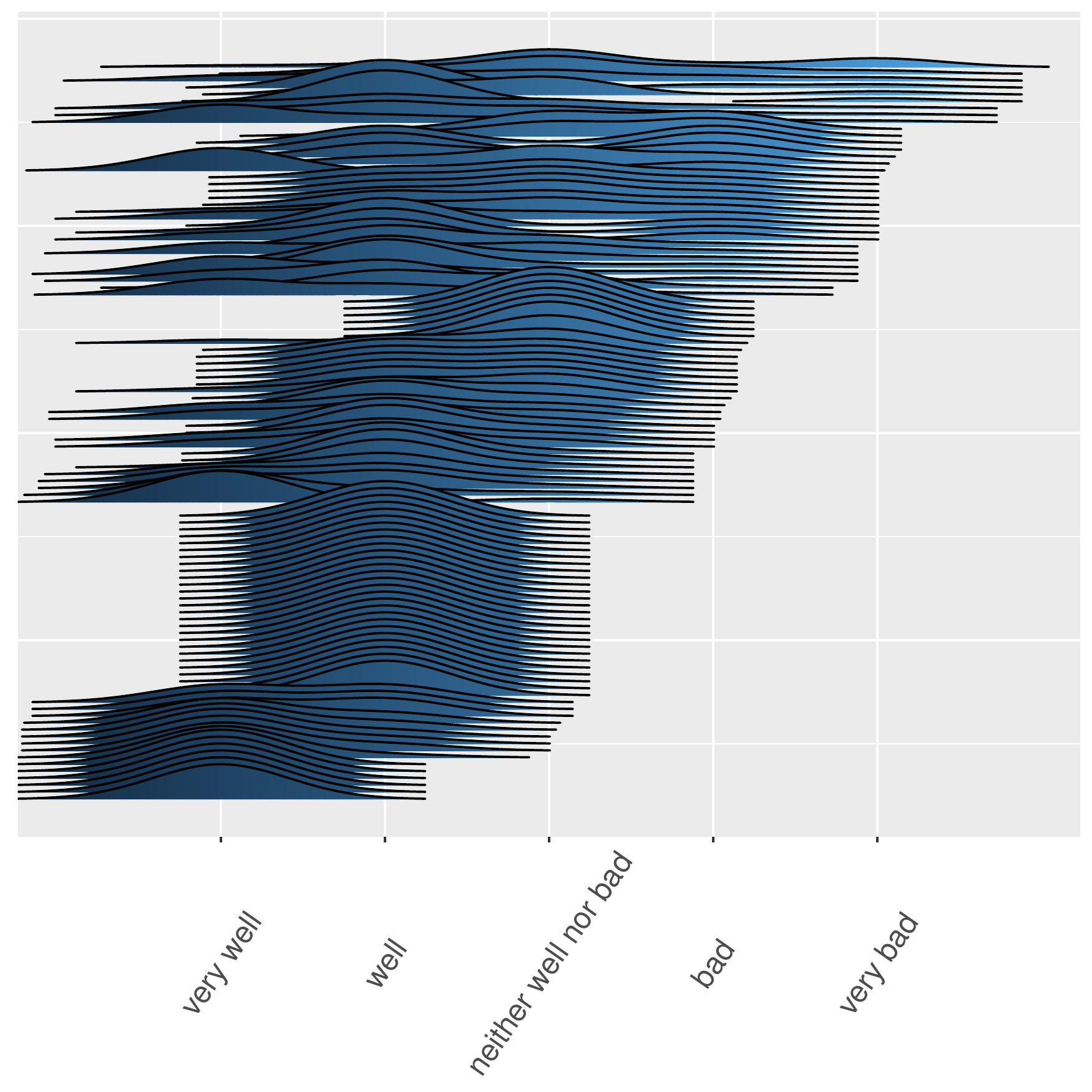}
	\caption{Distribution of structure quality of meta-models per participant}\label{fig:mm-structure-qual}
\end{figure}

Participants agreed that the vast majority of meta-models they transform are well structured ($x_{20}$).
This means there is little to no additional burden put onto development solely due to unfavourably structured meta-models.
The distribution of structure assessment per participant is shown in \Cref{fig:mm-structure-qual}.

\begin{figure}
	\includegraphics[width=\linewidth]{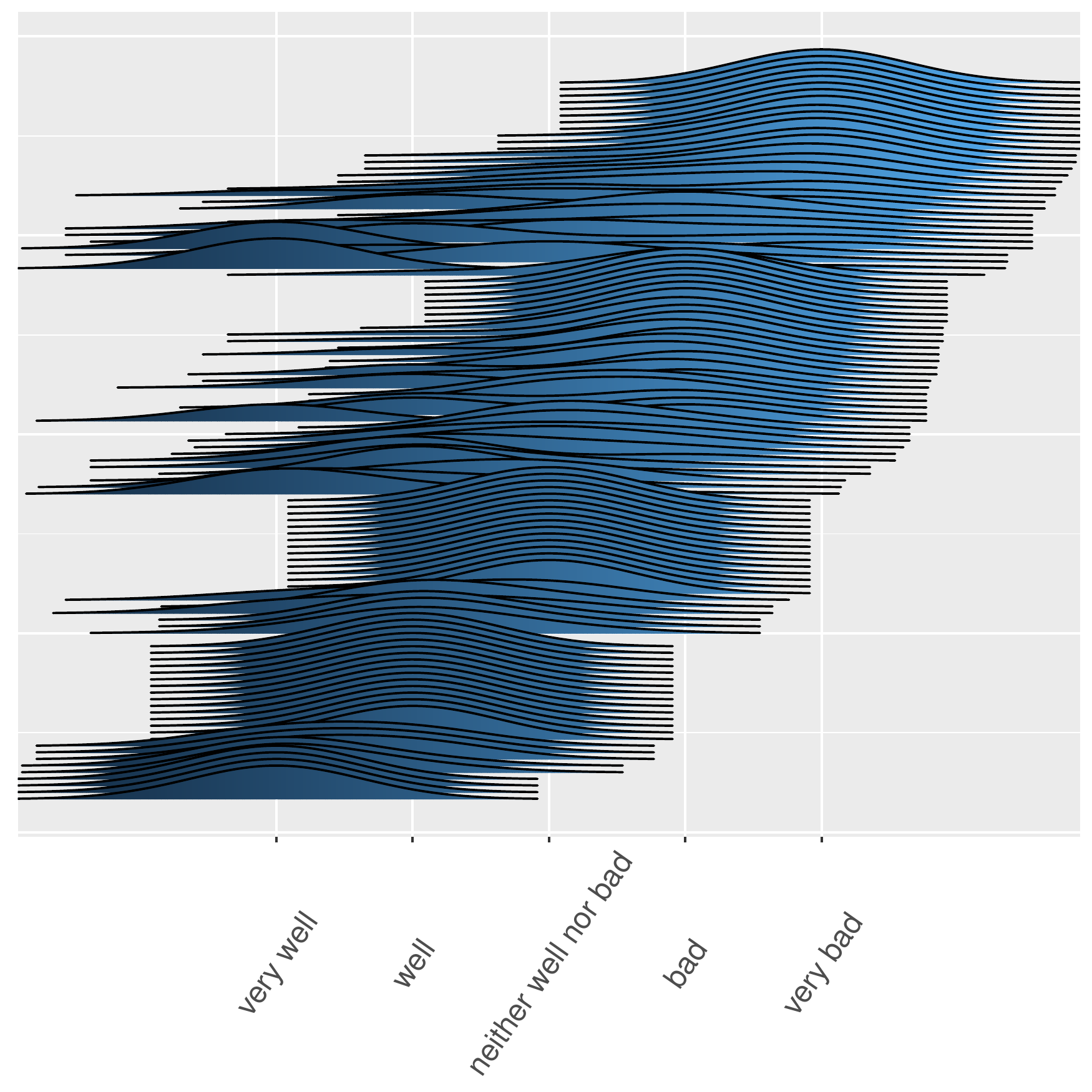}
	\caption{Distribution of meta-model documentation quality per participant}\label{fig:mm-docu-qual}
\end{figure}

The situation is different with documentation ($x_{21}$).
Most participants stated that they have experience with badly or even very baldy documented meta-models \Cref{fig:mm-docu-qual}.
For many participants, this constitutes the majority of meta-models they work with.

\section{Results}
\label{sec:results}

In this section, we present the results of our analysis of the questionnaire responses using universal structure modelling structured around the research questions \textbf{RQ1-4}.
The quantitative results for all influences between MTL capabilities and MTL properties are shown in \Cref{tbl:influences} in \Cref{apdx:results}.
On the x-axis the different \textit{MTL Properties} are shown.
On the y-axis the \textit{MTL Capabilities} are shown.
The first number in a cell describes the average simulated effect.
The second number describes the overall explained absolute deviation.
The third number shows the significance value.
A significance value lower or equal to 0.01* (the chosen significance level) is indicated with one asterisks.
The effect strengths of moderation effects can be found in \Cref{tbl:moderation:bx,tbl:moderation:exp,tbl:moderation:explang,tbl:moderation:inc,tbl:moderation:io,tbl:moderation:lang,tbl:moderation:mms,tbl:moderation:mmsanity,tbl:moderation:ms,tbl:moderation:ts} in \Cref{apdx:results}.
Each table describes the moderation effect of one of the moderating factors on all influences between MTL Capabilities and MTL Properties.

The rest of this section presents our results in context of the four research questions.
We focus on the most salient influences that we deem interesting for the respective research question.
Detailed interpretation and discussion of the implications of the presented results are done in \Cref{sec:discussion}.

\subsection{RQ1: Which of the hypothesised interdependencies withstands a test of significance? \& RQ4: What additional interdependencies arise from the analysis that were not initially hypothesised?}
\label{sec:results:rq14}

Our first research question is aimed at evaluating the accuracy of the structure model developed in the previous study~\parencite{Hoeppner2022}.
We do so by subjecting all hypothesised influences to a significance test during analysis.
The significance test can also be used to directly gain insights into interdependencies missed in the initial model.
Thus we discuss both the rejection of previously hypothesised influences as well as the extension of the model through newly discovered significant interdependencies in this section.

Most initially hypothesised influences withstand the test of significance but there are several exceptions.
Most notably all but one(\textit{Maintainability}) of the hypothesised influences of \textit{Bidirectionality} functionality of MTLs have to be rejected.
This means that from our results we can not conclude that the presence of \textit{Bidirectionality} functionality in a language changes how people perceive the \textit{Comprehensibility}, \textit{Ease of Writing} \textit{Expressiveness}, \textit{Tool Support}, \textit{Productivity} and \textit{Reusability} of the language.

Similarly, half of the influences on \textit{Ease of Writing} and \textit{Expressiveness} are also rejected.
This means that the presence of \textit{Bidirectionality}, \textit{Incrementality}, \textit{Model Management} and \textit{Model Traversal} functionality do not change how people perceive the \textit{Ease of Writing} transformations with a language.
And that the presence of \textit{Bidirectionality}, \textit{Incrementality}, \textit{Model Navigation}, \textit{Model Traversal} and \textit{Reuse} functionality do not change how people perceive the \textit{Expressiveness} of a language.

On a more positive note.
The hypothesised influences on \textit{Comprehensibility} are confirmed (apart from the one exerted by \textit{Bidirectionality}).
The same goes for \textit{Productivity} and \textit{Reusability}.

We also found that the perceived quality in \textit{Tool Support} and \textit{Maintainability} are influenced by most of the \textit{MTL Properties}.
A result that was not apparent from previous interview study.
\textit{Tool support} was hypothesised to be influenced only by the chosen language and \textit{Maintainability} only by \textit{Bidirectionality} and \textit{Mapping} functionality.
Our results however show, that the perceived \textit{Maintainability} of transformations written in a language is influenced by all MTL functionality considered in this study with the exception of \textit{Incrementality}.

Moreover, several additional influences on \textit{Productivity} and \textit{Reusability} were also discovered.
The perceived \textit{Productivity} and \textit{Reusability} of transformations in a language are influenced by \textit{Mappings}, \textit{Model Management}, \textit{Model Navigation}, \textit{Model Traversal}, \textit{Pattern Matching}, \textit{Reuse Mechanisms} and \textit{Tracability} functionality.

Regarding the moderating effects, our findings suggest that a nuanced view is warranted.
The hypothesis that context moderates all influences on an MTL Property still holds but the strength of the moderation effects varies greatly.

As hypothesised, we are able to observe that \textit{Comprehensibility} and \textit{Ease of Writing} are the two properties moderated by the most context variables.
But the moderation is only significant for a hand full of influences on these properties.
This can be seen e.g. in the moderation effects of \textit{Meta-Model Size} on influences on \textit{Comprehensibility} depicted in \Cref{tbl:moderation:ms} in \Cref{apdx:results}.
Changes in the \textit{Meta-model sizes} participants worked with had next to no effect on how their usage of \textit{Bidirectionality} functionality affected their view on the \textit{Comprehensibility} of transformations.
The impact on the influence of \textit{Model Management} on \textit{Comprehensibility} is orders of magnitudes higher.

Another observation that stands out is the impact of \textit{Language Choice} and \textit{Language Experience}.
The moderation effects of both variables are negligible or even 0 for all influences.
We believe this is due to the large number of languages considered in this study.
It makes analysing the effects of choosing one of the languages difficult.

Overall the results for research questions \textbf{RQ1 \& 4} suggest that our initial structure model contains many relevant interdependencies but several more have to be considered as well.
We do have to reject several direct influences due to low significance and moderation effects have to be considered on a per influence basis instead of being generalised for each \textit{MTL Property}.

\subsection{RQ2: How strong are the influences of model transformation language capabilities on the properties thereof?}

Our second research question is intended to provide numbers that can help to identify the most important factors to consider when evaluating the advantages and disadvantages of model transformation languages empirically.
We do this by considering both the average simulated effect of influences calculated by NEUSREL as well as the overall explained absolute deviation of influences compared to each other.
As explained earlier in this section all numbers can be found in \Cref{tbl:influences}.

\begingroup
\renewcommand{\arraystretch}{1.5}
\begin{sidewaystable*}
	\caption{Average simulated effect, overall explained absolute deviation and significance of direct influences}\label{tbl:influences}
	\begin{tabularx}{\textwidth}{>{\raggedright\arraybackslash\hspace{0pt}}X*{7}{>{\centering\arraybackslash\hspace{0pt}}X}}
		\toprule
		& Comprehensibility       & Ease of Writing        & Expressiveness         & Tool Support            & Maintainability        & Productivity         & Reusability         \\
		\midrule
		Bidirectionality   & -3.2e-07/ 3.3e-07/ 1    & 6.6e-08/ 9.2e-05/ 1    & 1.5e-09/ 4.2e-06/ 1    & -1.5e-05/ 2.0e-05/ 1    & -0.06/ 0.05/ 0.1*      & -8.3e-06/ 8.9e-07/ 1 & 1.4e-06/ 5.5e-06/ 1 \\
		Incrementality     & 0.02/ 0.01/ 0.01*       & -3.5e-07/ 0.0003/ 1    & 7.4e-07/ 0.002/ 1      & 0.0002/ 0.0002/ 1       & 0.0004/ 1.7e-05/ 1     & -0.0001/ 6.5e-07/ 1  & 0.02/ 0.005/ 0.05   \\
		Mappings           & 0.03/ 0.01/ 0.01*       & -1.8e-05/ 0.01/ 0.01*  & -6.8e-06/ 0.01/ 0.01*  & -0.0116/ 0.01/ 0.01*    & -0.000793/ 0.003/ 1    & -0.005/ 0.04/ 0.01*  & -0.02/ 0.01/ 0.01*  \\
		Model Management   & 0.03/ 0.1/ 0.01*        & 4.4e-05/ 0.003/ 1      & 0.0006/ 0.006/ 0.01*   & 0.02/ 0.07/ 0.01*       & 0.03/ 0.04/ 0.01*      & -0.0005/ 0.05/ 0.01* & 0.06/ 0.03/ 0.01*   \\
		Model Navigation   & -0.01/ 0.03/ 0.01*      & -2.3e-05/ 0.01/ 0.01*  & 5.9e-05/ 0.005/ 0.05   & 0.06/ 0.2/ 0.01*        & -0.004/ 0.05/ 0.01*    & 0.05/ 0.07/ 0.01*    & -0.08/ 0.08/ 0.01*  \\
		Model Traversal    & 0.008/ 0.009/ 0.01*     & 2.1e-07/ 0.002/ 1      & -9.4e-05/ 0.0005/ 1    & -0.09/ 0.2/ 0.01*       & 0.07/ 0.1/ 0.01*       & -0.003/ 0.03/ 0.01*  & 0.007/ 0.01/ 0.01*  \\
		Pattern Matching   & 0.05/ 0.07/ 0.01*       & -8.4e-05/ 0.01/ 0.01*  & -0.0001/ 0.006/ 0.01*  & 0.05/ 0.06/ 0.01*       & -0.04/ 0.08/ 0.01*     & 0.005/ 0.1/ 0.01*    & -0.06/ 0.06/ 0.01*  \\
		Reuse Mechanisms   & 0.1/ 0.08/ 0.01*        & -7.8e-05/ 0.008/ 0.01* & -0.0002/ 0.002/ 1      & 0.1/ 0.1/ 0.01*         & 0.1/ 0.2/ 0.01*        & 0.1/ 0.2/ 0.01*      & 0.2/ 0.1/ 0.01*     \\
		Traceability       & 0.29/ 0.12/ 0.01*       & 0.002/ 0.02/ 0.01*     & -2.1e-05/ 0.007/ 0.01* & -0.05/ 0.2/ 0.01*       & -0.02/ 0.1/ 0.01*      & 0.04/ 0.05/ 0.01*    & 0.08/ 0.09/ 0.01*   \\
		\bottomrule
	\end{tabularx}
	Please note that the significance values obtained through the NEUSREL tool may exhibit reduced accuracy compared to standard approaches due to the bootstrapping method used for their estimation.
\end{sidewaystable*}
\endgroup

Overall the effects identified in our analysis are lower than anticipated.
They range from 0.29 down till 6.5e-8.
We expected some effects to be low, mainly those from non significant interdependencies, but the fact that even significant effects are in the order of 0.01 is surprising.
We assume this stems from the large number of variables that are involved and the overall complexity of the matter under investigation.
Nonetheless we believe there are meaningful insights that can be drawn when comparing the influences for each \textit{MTL Property} with each other.

Of the influences hypothesised from our previous interview study \textit{Traceability} is the most impactful \textit{MTL Capability}.
Its usage exerts the highest influence on perceived \textit{Comprehensibility} with $0.29$.
Similarly it has the highest influence for \textit{Ease of Writing} though with a value of $0.0021$ the effect is small.
We were, however, already able to show empirical evidence that MTLs utilising automatic trace handling provide clear advantages for writing transformations compared to GPLs~\parencite{Hoeppner2021}.

For the properties \textit{Tool Support}, \textit{Maintainability} and \textit{Productivity} the availability of \textit{Reuse Mechanisms} seems to be the strongest driving factor with an average simulated effect of $0.1$, $0.1$, $0.1$ and $0.2$, respectively.
No other factor has an ASE or effect strength as high as \textit{Reuse Mechanisms} for these properties.
This result is surprising as the influences were not raised even once during our interview study.

Overall, automatic tracing and reuse mechanisms appear to be the most influential factors for MTL properties.
This suggests to us two main pathways for further research.
First, to improve model transformation languages more research should be devoted to developing effective ways to reuse transformations or parts of transformations.
From our experience, current mechanism are hard to use and are especially unsuited for different use-cases.
Secondly, the first area to address for improved adoption of model transformation concepts in general purpose languages should be the development of mechanisms for automatic trace handling.

\subsection{RQ3: How strong are moderation effects expressed by the contextual factors \textit{use-case}, \textit{skills \& experience} and \textit{MTL choice}?}

As expressed in \Cref{sec:results:rq14} the results of our analysis suggest that a more nuanced view of moderation effects is warranted.
In this section we go into detail on these nuances.

As hypothesised the size of meta-models moderates the influences on \textit{Comprehensibility}.
The moderation strength differs greatly between the different causing factors though.
For example, \textit{Meta-model size} exerts the strongest moderation on the influence of \textit{Model Management} onto \textit{Comprehensibility} with $0.14$.
All other moderation effects are far lower.
The second highest moderation effect, the moderation of \textit{Meta-model size} on the influence of \textit{Traceability} on \textit{Comprehensibility}, is about half es strong ($0.0778$) and the lowest, the moderation of \textit{Meta-model size} on the influence of \textit{Bidirectionality} functionality on \textit{Comprehensibility}, is only $0.0009$.
The moderations make sense intuitively as larger meta-models would make implementing these tasks manually more labour intensive and thus clutter the code unnecessarily.

\textit{Model size} exerts similar moderation effects as meta-model size.
Its strongest moderation effect is also on the influence of \textit{Model Management} on \textit{Comprehensibility} ($0.36$).
Moreover, \textit{Model size} also strongly moderates the influence of \textit{Traceability} functionality on the \textit{Ease of Writing} transformations ($0.17$).
Most other moderation effects of \textit{Model size} are far lower than $0.1$.

\textit{Transformation size} seems to be the most relevant moderating factors across the board.
It has many noteworthy moderation effects on all influences of \textit{MTL Capabilities} on \textit{Tool Support}, none being less than $0.16$, and \textit{Productivity}, most being above $0.12$.
We assume this is because the larger transformations get, the more reliant developers are on tooling and abstractions that reduce the development effort.

Another interesting effect we found is, that \textit{developer experience} moderates the influence of many of the domain specific abstractions, e.g. \textit{Mappings} and \textit{Model Traversal}, on \textit{Productivity}.
This makes sense because these specific features often break with how developers are used to develop programs and thus need practice to use them effectively.

The \textit{semantic gap between input and output} meta-models exerts its moderation strongest on the influences on \textit{Maintainability}.
Most notable are the moderations on the influences of \textit{Model Traversal} ($0.194$), \textit{Pattern Matching} ($0.239$) and \textit{Reuse Mechanisms} ($0.237$).

Lastly, there is a strong moderation effect of the \textit{meta-model sanity} onto the influence of \textit{Model Management} facilities and \textit{Bidirectionality} on \textit{Comprehensibility}.
Both being about $0.2$.
This makes sense as badly structured or poorly documented meta-models are harder to handle and thus the tasks revolving around working with the structure are most influenced by that.


Overall the size of transformations is in our opinion the most relevant moderating variable.
The assumption on the relevance of language choice could however not be confirmed.
This is most likely due to the large amount of languages each participant has had experience with which weakens the ability to elicit the effect of differences of language choice between participants.
\section{Discussion}
\label{sec:discussion}

The results of our analysis provide useful insights for research on model transformation languages.
In this section, we discuss the implications of our results for evaluation and development of MTLs.
Additionally, we provide a critical evaluation of our methodology with regards to the goals of this study.

\subsection{Implications of results}

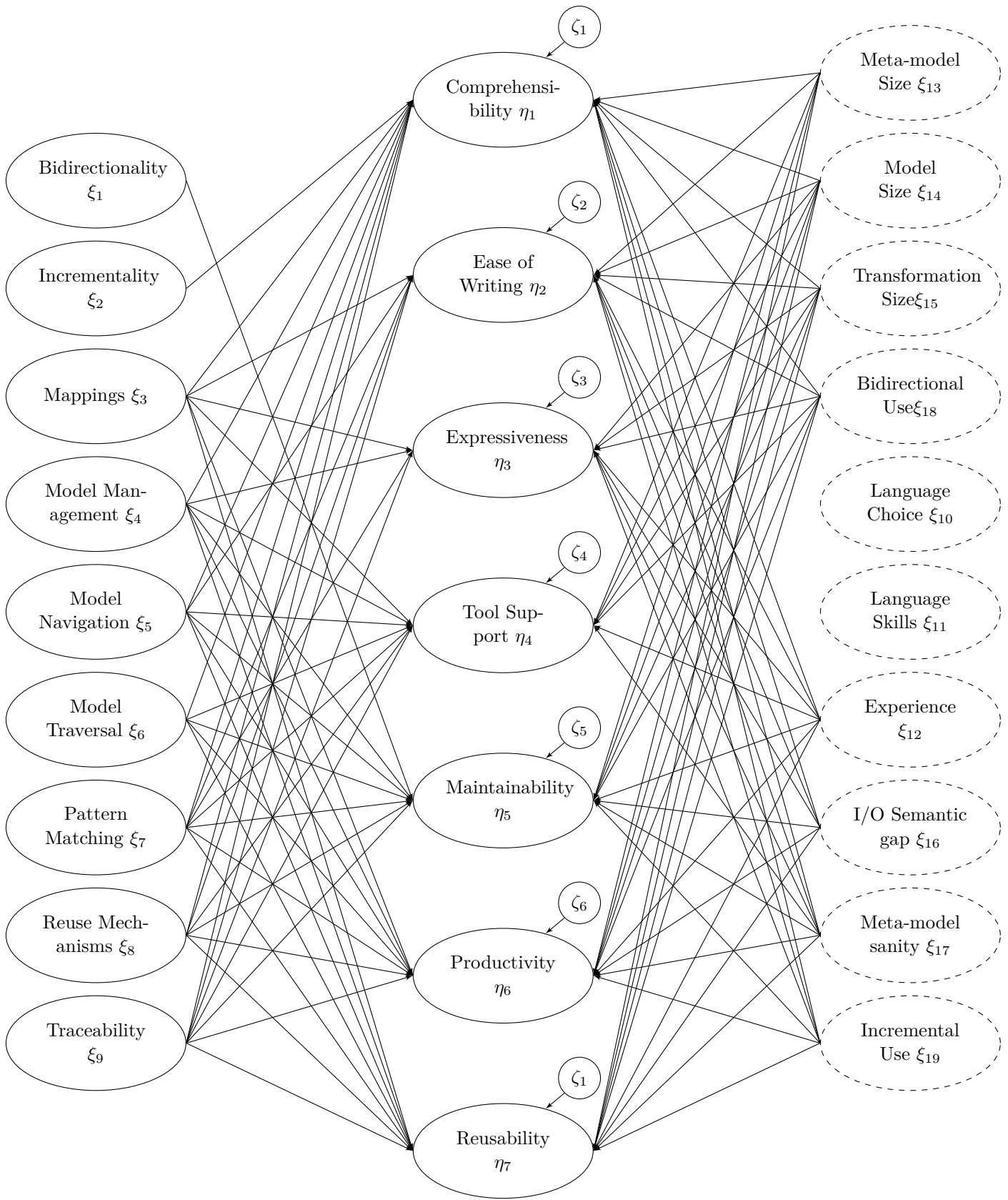
\begin{figure*}
	\begin{tikzpicture}
		\node[factor] (factor-bx) at (0,-2) {Bidirectionality $\xi_1$};
		\node[factor,below of=factor-bx] (factor-inc) {Incrementality $\xi_2$};
		\node[factor,below of=factor-inc] (factor-map) {Mappings $\xi_3$};
		\node[factor,below of=factor-map] (factor-man) {Model Management $\xi_4$};
		\node[factor,below of=factor-man] (factor-nav) {Model Navigation $\xi_5$};
		\node[factor,below of=factor-nav] (factor-trv) {Model Traversal $\xi_6$};
		\node[factor,below of=factor-trv] (factor-pm) {Pattern Matching $\xi_7$};
		\node[factor,below of=factor-pm] (factor-rm) {Reuse Mechanisms $\xi_8$};
		\node[factor,below of=factor-rm] (factor-trc) {Traceability $\xi_9$};
		
		\node[mod-factor] (mfactor-mms) at (15,0) {Meta-model Size $\xi_{13}$};
		\node[mod-factor,below of=mfactor-mms] (mfactor-ms) {Model Size $\xi_{14}$};
		\node[mod-factor,below of=mfactor-ms] (mfactor-ts) {Transformation Size$\xi_{15}$};
		\node[mod-factor,below of=mfactor-ts] (mfactor-bus) {Bidirectional Use$\xi_{18}$};
		\node[mod-factor,below of=mfactor-bus] (mfactor-mtl) {Language Choice $\xi_{10}$};
		\node[mod-factor,below of=mfactor-mtl] (mfactor-lsk) {Language Skills $\xi_{11}$};
		\node[mod-factor,below of=mfactor-lsk] (mfactor-exp) {Experience $\xi_{12}$};
		\node[mod-factor,below of=mfactor-exp] (mfactor-io) {I/O Semantic gap $\xi_{16}$};
		\node[mod-factor,below of=mfactor-io] (mfactor-msa) {Meta-model sanity $\xi_{17}$};
		\node[mod-factor,below of=mfactor-msa] (mfactor-ius) {Incremental Use $\xi_{19}$};
		
		\node[property] (prop-comp) at (7.5,-0.5) {Comprehensi-bility $\eta_1$};
		
		\node[error, above of=prop-comp] (comp-error) {$\zeta_1$};
		\path[l] (comp-error.south west) -- (prop-comp.45);
		
		\node[property,below of=prop-comp] (prop-eow) {Ease of Writing $\eta_2$};
		
		\node[error, above of=prop-eow] (eow-error) {$\zeta_2$};
		\path[l] (eow-error.south west) -- (prop-eow.45);
		
		\node[property,below of=prop-eow] (prop-ex) {Expressiveness $\eta_3$};
		
		\node[error, above of=prop-ex] (exp-error) {$\zeta_3$};
		\path[l] (exp-error.south west) -- (prop-ex.45);
		
		\node[property,below of=prop-ex] (prop-ts) {Tool Support $\eta_4$};
		
		\node[error, above of=prop-ts] (ts-error) {$\zeta_4$};
		\path[l] (ts-error.south west) -- (prop-ts.45);
		
		\node[property,below of=prop-ts] (prop-mtb) {Maintainability $\eta_5$};
		
		\node[error, above of=prop-mtb] (mtb-error) {$\zeta_5$};
		\path[l] (mtb-error.south west) -- (prop-mtb.45);
		
		\node[property,below of=prop-mtb] (prop-pro) {Productivity $\eta_6$};
		
		\node[error, above of=prop-pro] (pro-error) {$\zeta_6$};
		\path[l] (pro-error.south west) -- (prop-pro.45);
		
		\node[property,below of=prop-pro] (prop-reu) {Reusability $\eta_7$};
		
		\node[error, above of=prop-reu] (reu-error) {$\zeta_1$};
		\path[l] (reu-error.south west) -- (prop-reu.45);

		
		\path[l] (factor-bx.east) -- (prop-mtb.west);

		\path[l] (factor-inc.east) -- (prop-comp.west);
		
		\path[l] (factor-map.east) -- (prop-comp.west);
		\path[l] (factor-map.east) -- (prop-eow.west);
		\path[l] (factor-map.east) -- (prop-ex.west);
		\path[l] (factor-map.east) -- (prop-ts.west);
		\path[l] (factor-map.east) -- (prop-pro.west);
		\path[l] (factor-map.east) -- (prop-reu.west);
		
		\path[l] (factor-man.east) -- (prop-comp.west);
		\path[l] (factor-man.east) -- (prop-ex.west);
		\path[l] (factor-man.east) -- (prop-ts.west);
		\path[l] (factor-man.east) -- (prop-mtb.west);
		\path[l] (factor-man.east) -- (prop-pro.west);
		\path[l] (factor-man.east) -- (prop-reu.west);
		
		\path[l] (factor-nav.east) -- (prop-comp.west);
		\path[l] (factor-nav.east) -- (prop-eow.west);
		\path[l] (factor-nav.east) -- (prop-ts.west);
		\path[l] (factor-nav.east) -- (prop-mtb.west);
		\path[l] (factor-nav.east) -- (prop-pro.west);
		\path[l] (factor-nav.east) -- (prop-reu.west);
		
		\path[l] (factor-trv.east) -- (prop-comp.west);
		\path[l] (factor-trv.east) -- (prop-ts.west);
		\path[l] (factor-trv.east) -- (prop-mtb.west);
		\path[l] (factor-trv.east) -- (prop-pro.west);
		\path[l] (factor-trv.east) -- (prop-reu.west);
		
		\path[l] (factor-pm.east) -- (prop-comp.west);
		\path[l] (factor-pm.east) -- (prop-eow.west);
		\path[l] (factor-pm.east) -- (prop-ex.west);
		\path[l] (factor-pm.east) -- (prop-ts.west);
		\path[l] (factor-pm.east) -- (prop-mtb.west);
		\path[l] (factor-pm.east) -- (prop-pro.west);
		\path[l] (factor-pm.east) -- (prop-reu.west);
		
		\path[l] (factor-rm.east) -- (prop-comp.west);
		\path[l] (factor-rm.east) -- (prop-eow.west);
		\path[l] (factor-rm.east) -- (prop-ts.west);
		\path[l] (factor-rm.east) -- (prop-mtb.west);
		\path[l] (factor-rm.east) -- (prop-pro.west);
		\path[l] (factor-rm.east) -- (prop-reu.west);
		
		\path[l] (factor-trc.east) -- (prop-comp.west);
		\path[l] (factor-trc.east) -- (prop-eow.west);
		\path[l] (factor-trc.east) -- (prop-ex.west);
		\path[l] (factor-trc.east) -- (prop-ts.west);
		\path[l] (factor-trc.east) -- (prop-mtb.west);
		\path[l] (factor-trc.east) -- (prop-pro.west);
		\path[l] (factor-trc.east) -- (prop-reu.west);

		\path[l] (mfactor-mms.west) -- (prop-comp.east);
		\path[l] (mfactor-mms.west) -- (prop-eow.east);
		\path[l] (mfactor-mms.west) -- (prop-ts.east);
		\path[l] (mfactor-mms.west) -- (prop-mtb.east);
		\path[l] (mfactor-mms.west) -- (prop-pro.east);
		\path[l] (mfactor-mms.west) -- (prop-reu.east);
		
		\path[l] (mfactor-ms.west) -- (prop-comp.east);
		\path[l] (mfactor-ms.west) -- (prop-eow.east);
		\path[l] (mfactor-ms.west) -- (prop-ex.east);
		\path[l] (mfactor-ms.west) -- (prop-ts.east);
		\path[l] (mfactor-ms.west) -- (prop-mtb.east);
		\path[l] (mfactor-ms.west) -- (prop-pro.east);
		\path[l] (mfactor-ms.west) -- (prop-reu.east);
		
		\path[l] (mfactor-ts.west) -- (prop-comp.east);
		\path[l] (mfactor-ts.west) -- (prop-eow.east);
		\path[l] (mfactor-ts.west) -- (prop-ex.east);
		\path[l] (mfactor-ts.west) -- (prop-ts.east);
		\path[l] (mfactor-ts.west) -- (prop-mtb.east);
		\path[l] (mfactor-ts.west) -- (prop-pro.east);
		\path[l] (mfactor-ts.west) -- (prop-reu.east);
		
		\path[l] (mfactor-bus.west) -- (prop-comp.east);
		\path[l] (mfactor-bus.west) -- (prop-eow.east);
		\path[l] (mfactor-bus.west) -- (prop-ex.east);
		\path[l] (mfactor-bus.west) -- (prop-ts.east);
		\path[l] (mfactor-bus.west) -- (prop-mtb.east);
		\path[l] (mfactor-bus.west) -- (prop-pro.east);
		\path[l] (mfactor-bus.west) -- (prop-reu.east);
		
		
		
		\path[l] (mfactor-exp.west) -- (prop-comp.east);
		\path[l] (mfactor-exp.west) -- (prop-eow.east);
		\path[l] (mfactor-exp.west) -- (prop-ex.east);
		\path[l] (mfactor-exp.west) -- (prop-ts.east);
		\path[l] (mfactor-exp.west) -- (prop-mtb.east);
		\path[l] (mfactor-exp.west) -- (prop-pro.east);
		\path[l] (mfactor-exp.west) -- (prop-reu.east);
		
		\path[l] (mfactor-io.west) -- (prop-comp.east);
		\path[l] (mfactor-io.west) -- (prop-eow.east);
		\path[l] (mfactor-io.west) -- (prop-ex.east);
		\path[l] (mfactor-io.west) -- (prop-mtb.east);
		\path[l] (mfactor-io.west) -- (prop-pro.east);
		\path[l] (mfactor-io.west) -- (prop-reu.east);
		
		\path[l] (mfactor-msa.west) -- (prop-comp.east);
		\path[l] (mfactor-msa.west) -- (prop-eow.east);
		\path[l] (mfactor-msa.west) -- (prop-ex.east);
		\path[l] (mfactor-msa.west) -- (prop-ts.east);
		\path[l] (mfactor-msa.west) -- (prop-mtb.east);
		\path[l] (mfactor-msa.west) -- (prop-pro.east);
		\path[l] (mfactor-msa.west) -- (prop-reu.east);
		
		\path[l] (mfactor-ius.west) -- (prop-comp.east);
		\path[l] (mfactor-ius.west) -- (prop-eow.east);
		\path[l] (mfactor-ius.west) -- (prop-ex.east);
		\path[l] (mfactor-ius.west) -- (prop-mtb.east);
		\path[l] (mfactor-ius.west) -- (prop-pro.east);
		\path[l] (mfactor-ius.west) -- (prop-reu.east);
		
	\end{tikzpicture}
	\caption{Structure model depicting the confirmed influence and moderation effects of factors on MTL properties.}
	\label{fig:final_structure_model}
\end{figure*}

The topic of influences on the quality properties of model transformation language is vastly complex, as reflected in the already large structure model which we set out to analyse.
While we were able to reject some of the hypothesised influences, our analysis also identified several new influences.
As a result, the structure model depicting the influences grew in complexity, further highlighting the need for comprehensive studies of the factors that influence MTL quality properties.
The updated structure model can be seen in \Cref{fig:final_structure_model}.
It contains 36 more interdependencies than the one we started our analysis with.

Our analysis produced a number of interesting observations that have important implications for further research.
In particular, we now discuss the implications for empirical evaluations.
Additionally, we highlight the implications of our results for further development of MTLs and domain-specific features thereof.

\subsubsection{Suggestions for further empirical evaluation studies}

\textit{Traceability} is one of the most important factors to consider when it comes to the development of model transformations.
This is because it has the strongest influence on the perceived quality of both the ease of writing and the comprehensibility of the resulting code.
It is crucial to consider scenarios where tracing is involved in order to properly evaluate the value of MTL abstractions for writing and comprehending transformations.
Additionally, it is important to evaluate scenarios where tracing is not necessary to understand the difference that MTL abstractions can make.
To truly understand the relevance of this feature, it is also important to assess how many real-world use cases require it.
By taking all of these factors into account, it is possible to gain a comprehensive understanding of the value of MTL abstractions for writing and comprehending transformations.

For evaluation of Maintainability, \textit{Reuse Mechanisms} as well as \textit{Model Traversal} functionality are important capabilities to consider.
We therefore believe that researchers focusing on such an evaluation must make sure to use transformations that utilise these capabilities.
Moreover, the most important context to consider is the semantic gap between input and output meta-models.
Empirical evaluations focusing on maintainability should therefore make sure to evaluate transformation cases with varying degrees of differences between input and output meta-models.
These studies should then analyse how much the effectiveness of MTLs and GPLs changes in light of the semantic gap between input and output.

When selecting transformations for evaluation, it is essential to consider their size.
Our results have shown that size has the most significant impact on the influence of other factors on properties.
Put differently, the larger the transformation, the more noticeable the effect of all capabilities will be.
As such, it is imperative to focus on large transformation use-cases when designing a study to evaluate MTLs.

\subsubsection{Suggestions on language development}

For us, the most surprising finding of this study is the importance of reuse functionality.
The quality attributes tool support, maintainability, productivity and reusability are all most influenced by it.
This is especially surprising because there was no indication of this in our interviews~\parencite{Hoeppner2022}.
We suppose this influence stems from the fact that reuse mechanisms allow for more abstraction and thus less code that can be developed and maintained more efficiently.

As a result we believe that more focus should be put on developing transformation specific reuse mechanisms.
We are aware that some languages, e.g. ATL, already provide general reuse mechanisms through concepts like inheritance.
However, these concepts are limited by the fact that they rely on the object-oriented nature of the involved models.
This means that they can only be used to define reusable code within transformations of a single meta-model.
Defining transformation behaviour that can be reused between different meta-models is not possible.
But this would be important to further reduce redundancy in transformation development.

As result, we believe, that development of reuse mechanisms tailored to MTs is important to focus on.
In order to stand out compared to the reuse mechanisms of GPLs, it may be valuable to explore ways to define and reuse common transformation patterns independently of meta-models.
Higher order transformations are sometimes used to allow reuse too~\parencite{Kusel2015}, but from our experience current implementations are too cumbersome to be used productively.
\textcite{Chechik2016} provide a number of suggestions for transformation specific reuse mechanisms but to the best of our knowledge there exist no implementations of their concepts.

\subsection{Interesting observations outside of USM}

When discussing model transformation languages, it is often stated that they are only demonstrated on `toy examples' that have little to no real world value.
This argumentation has for example been raised several times in our previous interview study~\parencite{Hoeppner2022}.
However, the demographic data collected in our study disputes this.

There are several participants that stated to have worked solely on small transformations with small meta-and input models.
But this group is opposed by a similarly large group of participants that have worked with huge transformations, dissimilar and large meta-models as well as large inputs.
From this we conclude, that there are large use-cases where model transformations and MTLs are applied but they rarely get described in publications.
It seems likely that such examples are not used for highlighting important aspects authors want to discuss due to the space describing such cases would take up.
However, we argue that it is paramount that such case-studies are published to diminish the cynicism that MTLs are only useful for small examples.

Another noteworthy observation based on the demographic data of our participants is that documentation pertaining meta-models is predominantly perceived as inadequate.
We believe that this is primarily due to the fact that many of meta-models stem from research projects that prioritize expeditious prototyping over the long-term viability of the artefacts.
Nonetheless, we are convinced that there is an urgent need to enhance the documentation surrounding model transformations.
This issue is not limited solely to the meta-models, but also extends to the languages that are known for their challenging learning curve because of lack of tutorials~\parencite{Hoeppner2022}.


\subsection{Critical Assessment of the used methodology}

The appeal of using structural equation modelling for analysing the responses to our survey was to have a method of analysis that can be used to investigate a complex hypothesis system in its entirety.
Moreover, analysis is straight forward after an initial setup due to the sophisticated tooling for this methodology.
Instead of presenting participants with a case that they should assess we also opted for querying them on their overall assessment of MTL quality attributes.
These design decisions have implications and ramifications that we discuss in this section.

First, the effects observed in our study are small.
We assume this stems from the intricate and large structure model and the comparatively small sample size.
As explained in \Cref{sec:methodology} it is suggested to have between 5 to 10 times as many participants as the largest number of parameters to be estimated in each structural equation.
In light of the newly discovered paths in our structure model, the 113 total participants are close to the minimum sample size required.
Moreover, because of the large number of influences we do expect the influence of a single factor to be much smaller than in structure models where only 2-3 factors are relevant.
The results therefore reinforce our assessment that it is a very complex topic.


We also ran into some difficulties when using NEUSREL to analyse our data.
The structure model was so large that sometimes the tool crashed during calculations.
The online tooling to set everything up was also painfully inefficient leading to more problems during setup like browser crashes.
It took us some trial and error to find a way to get everything set up and run the analysis without crashes.

We chose to execute a study based on our study design in hopes of producing a complete theory independent of the use case under consideration.
The results exhibit less effect strength but we believe them to be more externally valid.
Nonetheless, we think that several additional studies need to be conducted to confirm our results for different use-cases.
\section{Threats to validity}
\label{sec:threats}

Our study is carefully designed and follows standard procedures for this type of study.
There are, however still threats to validity that stem from design decisions and limitations.
In this section we discuss these threats.

\subsection{Internal Validity}
Internal validity is threatened by manual errors and biases of the involved researchers throughout the process.
	
The two activities where such errors and biases can be introduced are the subject selection and question creation.
The selection criteria for study subjects is designed in such a way, that no ambiguities exist during selection.
This prevents researcher bias.

The survey questions and answers to the questions pose another threat to internal validity.
We used neutral questions to prevent subconsciously influencing the opinions of research subjects.
We also provide explanations for ambiguous terms used in the survey.
However, there are several instances where we can not fully ensure that each participant interprets terms the same way.
The questions on quality properties of model transformation languages allow room for interpretation in that we do not provide a clear metric what terms such as `Very Comprehensible' or `Very Hard to write' mean.
Similarly, the questions on meta-model quality leave room for interpretation on the side of participants.
We opted for this limitation because there are no universal ways to quantify such estimates and because the subjective assessment is what we want to collect.
The reason for this is, that subjective experiences are the main driving factor for all discussions on development when people are the main subject.

To ensure overall understandability and prevent errors in the setup of the survey we used a pilot study.

\subsection{External Validity}
External validity is threatened by our subject sampling strategy and the limitations on the survey questions imposed by the complexity of the subject matter.
	
We utilise convenience sampling.
Convenience sampling can limit how representative the final group of interviewees is.
Since we do not know the target populations makeup, it is difficult to asses the extend of this problem.
	
Using research articles as a starting point introduces a bias towards researchers.
There is little potential to mitigate this problem during the study design, because there exists no systematic way to find industry users.
	
Due to the complexity and abstractness of the concepts under investigation, a measurement via reflective of formative indicators is not possible.
Instead we use single item questions.
We further assume that positive and negative effects of a feature are more prominent if the feature is used more frequently.
This can have a negative effect on the external validity of our results.
However, we consciously decided for these limitations to be able to create a study that concerns itself with all factors and influences at once.

\subsection{Construct Validity}
Construct validity is threatened by inappropriate methods used for the study.
	
Using the results of online surveys as input for structural equation modelling techniques is common practice in market research~\parencite{Weiber2021}.
It is less common in computer science.
However, we argue that for the purpose of our study it is an appropriate methodology.
This is because the goal of extracting influence strengths and moderation effects of factors on different properties aligns with the goals of market research studies that employ structural equation modelling.

\subsection{Conclusion Validity}
Conclusion validity is mainly threatened by biases of our survey participants.
	
It is possible that people who do research on model transformation languages or use them for a long time are more likely to see them in a positive light.
As such there is the risk that too little experiences will be reported on in our survey.
However, this problem did not present itself in a previous study by us on the subject matter \parencite{Hoeppner2022}.
In fact researchers were far more critical in dealing with the subject.
As a result, there might be a slight positive bias in the survey responses, but we believe this to be negligible.
\section{Related Work}
\label{sec:rw}

There are numerous works that explore the possibilities gained through the usage of MTLs such as automatic parallelisation~\parencite{9146715,biermann2010lifting,benelallam2015distributed}, verification~\parencite{lano2015framework,Ko2015} or simply the application of difficult transformations~\parencite{10.1007/978-3-540-75209-7_30}.
There is, however, only a small amount of works trying to evaluate the languages to gain insights into where specific advantages or disadvantages associated with the use of MTLs originate from.
Several other works that can be related to our study also exist.
The related work is divided into studies focused on the investigation of properties of model transformation languages and empirical studies on model transformation languages.

\subsection{Studies on the Properties of Model Transformation Languages}

\textcite{Burgueno2019} conducted on a online survey and open discussion at the 12th edition of the International Conference on Model Transformations (ICMT'2019).
The goal of the survey was to identify reasons why developers decided to use or dismiss MTLs for writing transformations.
They also tried to gauge the communities sentiment on the future of model transformation languages.
At ICMT'2019, where the results of the survey were presented, they then held an open discussion on this topic and collected the responses of participants.
Their results show that MTLs have fallen in popularity.
They attribute this to 3 types of issues, technical issues, tooling issues and social issues, as well as the fact that GPLs have assimilated many ideas from MTLs.
The results of their study are a major driver in the motivation of our work.
While they identified issues and potential avenues for future research, their results are qualitative and broad which we try to improve upon with our study.

In a prior study of ours~\parencite{Goetz2020}, we conducted a structured literature review which forms the basis of much of our work since then.
The literature review aimed at extracting and categorising claims about the advantages and disadvantages of model transformation languages as well as the state of empirical evaluation thereof.
We searched over 4000 publication for this purpose and extracted 58 that directly claim properties of MTLs.
In total 137 claims were found and categorised into 15 quality properties of model transformation languages.
The results of the study show that little to no empirical studies to evaluate MTLs exist and that there is a severe lack of context and background information that further hinders their evaluation.

Lastly, there is our interview study~\parencite{Hoeppner2022} the data of which forms the basis for the reported study.
We interviewed 56 people on what they believe the most relevant factors are that facilitate or hamper their advantages for different quality properties identified in the prior literature review.
The interviews brought forth insights into factors from which the advantages and disadvantages of MTLs originate from as well as suggested a number of moderation effects on the effects of these factors.
These results for the data basis for this study.

\subsection{Empirical Studies on Model Transformation Languages}

\textcite{Hebig2018} report on a controlled experiment to evaluate how the use of different languages, namely ATL, QVT-O and Xtend affects the outcome of students solving several transformation tasks.
During the study student participants had to complete a series of three model transformation tasks.
One task was focused on comprehension, one task focused on modifying an existing transformation and one task required participants to develop a transformation from scratch.
The authors compared how the use of ATL, QVTo and Xtend affected the outcome of each of the tasks.
Unfortunately their results show no clear evidence of an advantage when using a model transformation language compared to Xtend.
However, they concede that the conditions under which the observations are made, were narrow.

We published a study on how much complexity stems from what parts of ATL transformations~\parencite{gotz2020investigating} and compared these results with data for transformations written in Java~\parencite{Hoeppner2021} to elicit advantageous features in ATL and to explore what use-cases justify the use of a general purpose language over a model transformation language.
In the study, the complexity of transformations written in ATL were compared to the same transformations written in Java SE5 and Java SE14 allowing for a comparison and historical perspective.
The Java transformations were translated from the ATL transformations using a predefined translation schema.
The results show that new language features in Java, like the Streams API, allow for significant improvement over older Java code, the relative amount of complexity aspects that ATL can hide stays the same between the two versions.

\textcite{Gerpheide2016} use a mixed method study consisting of expert interviews, a literature review and introspection, to formalize a quality model for the QVTo model transformation standard.
The quality model is validated using a survey and used to identify the necessity of quality tool support for developers.

We know of two study templates for evaluating model transformation languages that have been proposed but not yet used.
\textcite{Kramer2016} propose a template for a controlled experiment to evaluate comprehensibility of MTLs.
The template envisages using a questionnaire to evaluate the ability of participants to understand what presented transformation code does.
The influence of the language used for the transformation should then be measured by comparing the average number of correct answers and average time spent to fill out the questionnaire.
\textcite{Strueber2016} also propose a template for a controlled experiment.
The aim of the study is to evaluate the benefits and drawbacks of \textit{rule refinement} and \textit{variability-based rules} for reuse.
The quality of reusability is measured through measuring the comprehensibility as well as the changeability collected in bug-fixing and modification tasks.
\section{Conclusion}
\label{sec:conclusion}

Our study provides the first quantification of the importance of model transformation language capabilities for the perception of quality attributes by developers.
It once again highlight the complexity of the subject matter as the effect sizes of the influences are small and the final structure model grew in size.

As demonstrated by the amount of influences contained in the structure model many language capabilities need to be considered when designing empirical studies on MTLs.
The results however point towards Traceability and Reuse Mechanisms as the two most important MTL capabilities.
Moreover, the size of the transformations provides the strongest moderation effects to many of influences and is thus the most important context factor to consider.

Apart from implications for further empirical studies our results also point a clear picture for further language development.
Transformation specific reuse mechanisms should be the main focus as shown by their relevance for many development lifecycle focused quality attributes such as Maintainability and Productivity.


\section*{Declarations}
\textbf{Conflict of Interests} The authors have no competing interests to declare that are relevant to the content of this article.
%
\begingroup
\raggedright
\sloppy
\printbibliography

@String{Computer = "{IEEE} Computer" }

@String{Computing = "Computing" }

@String{Springer = "Springer-Verlag" }

@Article{fuchs2009using,
  author    = {Fuchs, Christoph and Diamantopoulos, Adamantios},
  date      = {2009},
  title     = {Using single-item measures for construct measurement in management research: Conceptual issues and application guidelines},
  number    = {2},
  pages     = {195},
  volume    = {69},
  journal   = {Die Betriebswirtschaft},
  publisher = {Schaeffer Poeschel Verlag},
  year      = {2009},
}

@Article{Sendall2003,
  author       = {S. {Sendall} and W. {Kozaczynski}},
  date         = {2003},
  journaltitle = {IEEE Software},
  title        = {Model transformation: the heart and soul of model-driven software development},
  doi          = {10.1109/MS.2003.1231150},
  year         = {2003},
}

@InCollection{metzger2005systematic,
  author    = {Metzger, Andreas},
  booktitle = {Model-driven Software Development},
  date      = {2005},
  title     = {A systematic look at model transformations},
  doi       = {10.1007/3-540-28554-7_2},
  pages     = {19--33},
  publisher = {Springer},
  year      = {2005},
}

@Article{Schmidt2006,
  author       = {Schmidt, Douglas},
  date         = {2006},
  journaltitle = {COMPUTER-IEEE COMPUTER SOCIETY},
  title        = {Guest Editor's Introduction: Model-Driven Engineering},
  doi          = {10.1109/MC.2006.58},
  year         = {2006},
}

@Article{Goetz2020,
  author       = {Götz, Stefan and Tichy, Matthias and Groner, Raffaela},
  date         = {2021},
  journaltitle = {Software and Systems Modeling},
  title        = {Claimed advantages and disadvantages of (dedicated) model transformation languages: a systematic literature review},
  doi          = {10.1007/s10270-020-00815-4},
  issn         = {1619-1374},
  number       = {2},
  pages        = {469--503},
  url          = {https://doi.org/10.1007/s10270-020-00815-4},
  volume       = {20},
  refid        = {Götz2021},
  year         = {2021},
}

@InProceedings{Hermans2009,
  author    = {Hermans, Felienne and Pinzger, Martin and van Deursen, Arie},
  booktitle = {Model Driven Engineering Languages and Systems},
  date      = {2009},
  title     = {Domain-Specific Languages in Practice: A User Study on the Success Factors},
  doi       = {10.1007/978-3-642-04425-0_33},
  isbn      = {978-3-642-04425-0},
  series    = {MODELS 2009},
  year      = {2009},
}

@InProceedings{Johannes2009,
  author    = {Johannes, Jendrik and Zschaler, Steffen and Fern{\'a}ndez, Miguel A. and Castillo, Antonio and Kolovos, Dimitrios S. and Paige, Richard F.},
  booktitle = {Model Driven Engineering Languages and Systems},
  date      = {2009},
  title     = {Abstracting Complex Languages through Transformation and Composition},
  doi       = {10.1007/978-3-642-04425-0_41},
  series    = {MODELS 2009},
  year      = {2009},
}

@Article{Burgueno2019,
  author       = {Loli Burgueño; Jordi Cabot and Sébastien Gérard},
  date         = {2019},
  journaltitle = {Journal of Object Technology},
  title        = {The Future of Model Transformation Languages: An Open Community Discussion},
  doi          = {10.5381/jot.2019.18.3.a7},
  year         = {2019},
}

@InProceedings{Hebig2018,
  author    = {Hebig, Regina and Seidl, Christoph and Berger, Thorsten and Pedersen, John Kook and W\k{a}sowski, Andrzej},
  booktitle = {Proceedings of the 2018 26th ACM Joint Meeting on European Software Engineering Conference and Symposium on the Foundations of Software Engineering},
  date      = {2018},
  title     = {Model Transformation Languages Under a Magnifying Glass: A Controlled Experiment with Xtend, ATL, and QVT},
  doi       = {10.1145/3236024.3236046},
  series    = {ESEC/FSE 2018},
  year      = {2018},
}

@Article{Hoeppner2021,
  author       = {Höppner, Stefan and Kehrer, Timo and Tichy, Matthias},
  date         = {2021},
  journaltitle = {Software and Systems Modeling},
  title        = {Contrasting dedicated model transformation languages versus general purpose languages: a historical perspective on ATL versus Java based on complexity and size},
  doi          = {10.1007/s10270-021-00937-3},
  issn         = {1619-1374},
  refid        = {Höppner2021},
  year         = {2021},
}

@Article{buckler2008identifying,
  author    = {Buckler, Frank and Hennig-Thurau, Thorsten},
  date      = {2008},
  title     = {Identifying hidden structures in marketing’s structural models through universal structure modeling},
  number    = {JRM 2},
  pages     = {47--66},
  volume    = {30},
  journal   = {Marketing ZFP},
  publisher = {Verlag Franz Vahlen},
  year      = {2008},
}

@Book{Weiber2021,
  author    = {Weiber, Rolf and M{\"u}hlhaus, Daniel},
  date      = {2021},
  title     = {Strukturgleichungsmodellierung: Eine anwendungsorientierte Einf{\"u}hrung in die Kausalanalyse mit Hilfe von AMOS, SmartPLS und SPSS},
  doi       = {10.1007/978-3-658-32660-9},
  edition   = {3},
  publisher = {Springer-Verlag},
  year      = {2021},
}

@Article{Mens2006,
  author       = {Tom Mens and Pieter Van Gorp},
  date         = {2006},
  journaltitle = {Electronic Notes in Theoretical Computer Science (GraMoT 2005)},
  title        = {A Taxonomy of Model Transformation},
  doi          = {10.1016/j.entcs.2005.10.021},
  year         = {2006},
}

@Article{kahani2019survey,
  author       = {Kahani, Nafiseh and Bagherzadeh, Mojtaba and Cordy, James R and Dingel, Juergen and Varr{\'o}, Daniel},
  date         = {2019},
  journaltitle = {Software \& Systems Modeling},
  title        = {Survey and classification of model transformation tools},
  doi          = {10.1007/s10270-018-0665-6},
  year         = {2019},
}

@Article{Czarnecki2006,
  author       = {K. {Czarnecki} and S. {Helsen}},
  date         = {2006},
  journaltitle = {IBM Systems Journal},
  title        = {Feature-based survey of model transformation approaches},
  doi          = {10.1147/sj.453.0621},
  year         = {2006},
}

@Article{9146715,
  author       = {Sanchez Cuadrado, Jesus and Burgueno, Loli and Wimmer, Manuel and Vallecillo, Antonio},
  date         = {2020},
  journaltitle = {IEEE Transactions on Software Engineering},
  title        = {Efficient execution of ATL model transformations using static analysis and parallelism},
  doi          = {10.1109/TSE.2020.3011388},
  pages        = {1-1},
  year         = {2020},
}

@Article{biermann2010lifting,
  author       = {Biermann, Enrico and Ermel, Claudia and Taentzer, Gabriele},
  date         = {2010},
  journaltitle = {Electronic Communications of the EASST},
  title        = {Lifting parallel graph transformation concepts to model transformation based on the eclipse modeling framework},
  volume       = {26},
  year         = {2010},
}

@InProceedings{benelallam2015distributed,
  author    = {Benelallam, Amine and G{\'o}mez, Abel and Tisi, Massimo and Cabot, Jordi},
  booktitle = {Proceedings of the 2015 ACM SIGPLAN International Conference on Software Language Engineering},
  date      = {2015},
  title     = {Distributed Model-to-model Transformation with ATL on MapReduce},
  pages     = {37--48},
  year      = {2015},
}

@Article{lano2015framework,
  author       = {Lano, Kevin and Clark, Tony and Kolahdouz-Rahimi, S},
  date         = {2015},
  journaltitle = {Formal Aspects of Computing},
  title        = {A framework for model transformation verification},
  number       = {1},
  pages        = {193--235},
  volume       = {27},
  publisher    = {Springer},
  year         = {2015},
}

@Article{Ko2015,
  author       = {Ko, Jong-Won and Chung, Kyung-Yong and Han, Jung-Soo},
  date         = {2015},
  journaltitle = {Multimedia Tools and Applications},
  title        = {Model transformation verification using similarity and graph comparison algorithm},
  doi          = {10.1007/s11042-013-1581-y},
  issn         = {1573-7721},
  number       = {20},
  pages        = {8907--8920},
  url          = {https://doi.org/10.1007/s11042-013-1581-y},
  volume       = {74},
  refid        = {Ko2015},
  year         = {2015},
}

@InProceedings{10.1007/978-3-540-75209-7_30,
  author    = {Anastasakis, Kyriakos and Bordbar, Behzad and Georg, Geri and Ray, Indrakshi},
  booktitle = {Model Driven Engineering Languages and Systems},
  date      = {2007},
  title     = {UML2Alloy: A Challenging Model Transformation},
  editor    = {Engels, Gregor and Opdyke, Bill and Schmidt, Douglas C. and Weil, Frank},
  isbn      = {978-3-540-75209-7},
  location  = {Berlin, Heidelberg},
  pages     = {436--450},
  publisher = {Springer Berlin Heidelberg},
  year      = {2007},
}

@Article{gotz2020investigating,
  author       = {G{\"o}tz, Stefan and Tichy, Matthias},
  date         = {2020},
  journaltitle = {J. Object Technol.},
  title        = {Investigating the Origins of Complexity and Expressiveness in ATL Transformations.},
  number       = {2},
  pages        = {12--1},
  volume       = {19},
  year         = {2020},
}

@Book{mooney1993bootstrapping,
  author    = {Mooney, Christopher Z and Mooney, Christopher F and Mooney, Christopher L and Duval, Robert D and Duvall, Robert},
  date      = {1993},
  title     = {Bootstrapping: A nonparametric approach to statistical inference},
  number    = {95},
  publisher = {sage},
  year      = {1993},
}

@Article{Hoeppner2022,
  author   = {Höppner, Stefan and Haas, Yves and Tichy, Matthias and Juhnke, Katharina},
  title    = {Advantages and disadvantages of (dedicated) model transformation languages},
  doi      = {10.1007/s10664-022-10194-7},
  issn     = {1573-7616},
  number   = {6},
  pages    = {159},
  volume   = {27},
  journal  = {Empirical Software Engineering},
  refid    = {Höppner2022},
  year     = {2022},
}

@Article{JOT:issue_2021_02/article5,
  author  = {Raffaela Groner and Katharina Juhnke and Stefan Götz and Matthias Tichy and Steffen Becker† and Vijayshree Vijayshree and Sebastian Frank},
  title   = {A Survey on the Relevance of the Performance of Model Transformations},
  doi     = {10.5381/jot.2021.20.2.a5},
  issn    = {1660-1769},
  note    = {OPEN REGULAR ISSUE},
  number  = {2},
  pages   = {2:1-27},
  url     = {http://www.jot.fm/contents/issue_2021_02/article5.html},
  volume  = {20},
  journal = {Journal of Object Technology},
  year    = {2021},
}

@Article{10.1145/3469888,
  author     = {Graziotin, Daniel and Lenberg, Per and Feldt, Robert and Wagner, Stefan},
  title      = {Psychometrics in Behavioral Software Engineering: A Methodological Introduction with Guidelines},
  doi        = {10.1145/3469888},
  number     = {1},
  volume     = {31},
  address    = {New York, NY, USA},
  articleno  = {7},
  issue_date = {January 2022},
  journal    = {ACM Trans. Softw. Eng. Methodol.},
  month      = {sep},
  numpages   = {36},
  publisher  = {Association for Computing Machinery},
  year       = {2021},
}

@Book{hinkel2016nmf,
  author    = {Hinkel, Georg},
  title     = {NMF: A Modeling Framework for the. NET Platform},
  publisher = {KIT},
  year      = {2016},
}

@Article{vanDeuersen2002,
  author   = {Van Deursen, Arie and Klint, Paul},
  title    = {Domain-specific language design requires feature descriptions},
  doi      = {10.2498/cit.2002.01.01},
  journal  = {Journal of Computing and Information Technology},
  year     = {2002},
}

@Article{Sprinkle2009,
  author   = {J. {Sprinkle} and M. {Mernik} and J. {Tolvanen} and D. {Spinellis}},
  title    = {Guest Editors' Introduction: What Kinds of Nails Need a Domain-Specific Hammer?},
  doi      = {10.1109/MS.2009.92},
  number   = {4},
  pages    = {15-18},
  volume   = {26},
  journal  = {IEEE Software},
  year     = {2009},
}

@Book{Kernighan1984,
  author    = {Brian W. Kernighan and Rob Pike},
  title     = {The Unix Programming Environment},
  isbn      = {0-13-937699-2},
  publisher = {Prentice Hall, Inc.},
  year      = {1984},
}

@Article{Raggett1999,
  author   = {Raggett, Dave and Le Hors, Arnaud and Jacobs, Ian and others},
  title    = {HTML 4.01 Specification},
  volume   = {24},
  journal  = {W3C recommendation},
  year     = {1999},
}

@Standard{SAEMobilus2004,
  author   = {SAEMobilus},
  title    = {Architecture Analysis and Design Language (AADL)},
  year     = {2004},
}

@Standard{OMG2001,
  author      = {OMG},
  institution = {Object Management Group},
  month       = jul,
  title       = {Model Driven Architecture (MDA), ormsc/2001-07-01},
  year        = {2001},
}

@InCollection{Brown2005,
  author    = {Brown, Alan W and Conallen, Jim and Tropeano, Dave},
  booktitle = {Model-Driven Software Development},
  title     = {Introduction: Models, modeling, and model-driven architecture (mda)},
  doi       = {10.1007/3-540-28554-7_1},
  pages     = {1--16},
  publisher = {Springer},
  year      = {2005},
}

@Article{Selic2003,
  author   = {B. {Selic}},
  title    = {The pragmatics of model-driven development},
  doi      = {10.1109/MS.2003.1231146},
  number   = {5},
  pages    = {19-25},
  volume   = {20},
  journal  = {IEEE Software},
  year     = {2003},
}

@Standard{OMG2016,
  author      = {OMG},
  institution = {Object Management Group},
  month       = apr,
  title       = {Meta Object Facility(MOF)},
  year        = {2002},
}

@Book{steinberg2008emf,
  author    = {Steinberg, Dave and Budinsky, Frank and Merks, Ed and Paternostro, Marcelo},
  title     = {{{EMF: eclipse modeling framework}}},
  publisher = {Pearson Education},
  year      = {2008},
}

@InProceedings{Arendt2010,
  author    = {Arendt, Thorsten and Biermann, Enrico and Jurack, Stefan and Krause, Christian and Taentzer, Gabriele},
  booktitle = {Model Driven Engineering Languages and Systems},
  title     = {Henshin: Advanced Concepts and Tools for In-Place EMF Model Transformations},
  doi       = {10.1007/978-3-642-16145-2_9},
  series    = {MODELS 2010},
  year      = {2010},
}

@InProceedings{Balogh2006,
  author    = {Balogh, Andr\'{a}s and Varr\'{o}, D\'{a}niel},
  booktitle = {Proceedings of the 2006 ACM Symposium on Applied Computing},
  title     = {Advanced Model Transformation Language Constructs in the VIATRA2 Framework},
  doi       = {10.1145/1141277.1141575},
  series    = {SAC '06},
  year      = {2006},
}

@InProceedings{Jouault2006,
  author    = {Jouault, Fr{\'e}d{\'e}ric and Allilaire, Freddy and B{\'e}zivin, Jean and Kurtev, Ivan and Valduriez, Patrick},
  booktitle = {Companion to the 21st ACM SIGPLAN Symposium on Object-oriented Programming Systems, Languages, and Applications},
  title     = {ATL: A QVT-like Transformation Language},
  doi       = {10.1145/1176617.1176691},
  series    = {OOPSLA '06},
  year      = {2006},
}

@InProceedings{P86,
  author    = {Kolovos, Dimitrios S. and Paige, Richard F. and Polack, Fiona A. C.},
  booktitle = {Theory and Practice of Model Transformations},
  title     = {The Epsilon Transformation Language},
  doi       = {10.1007/978-3-540-69927-9_4},
  series    = {ICMT 2008},
  year      = {2008},
}

@InProceedings{10.1007/978-3-642-38883-5_7,
  author    = {Horn, Tassilo},
  booktitle = {Theory and Practice of Model Transformations},
  title     = {Model Querying with FunnyQT},
  editor    = {Duddy, Keith and Kappel, Gerti},
  isbn      = {978-3-642-38883-5},
  pages     = {56--57},
  publisher = {Springer Berlin Heidelberg},
  address   = {Berlin, Heidelberg},
  year      = {2013},
}

@InProceedings{P94,
  author    = {George, Lars and Wider, Arif and Scheidgen, Markus},
  booktitle = {Theory and Practice of Model Transformations},
  title     = {Type-Safe Model Transformation Languages as Internal DSLs in Scala},
  doi       = {10.1007/978-3-642-30476-7_11},
  series    = {ICMT 2012},
  year      = {2012},
}

@Article{P23,
  author   = {Hinkel, Georg and Burger, Erik},
  title    = {Change propagation and bidirectionality in internal transformation DSLs},
  doi      = {10.1007/s10270-017-0617-6},
  journal  = {Software {\&} Systems Modeling},
  year     = {2019},
}

@InProceedings{Cuadrado2006,
  author    = {Cuadrado, Jes{\'u}s S{\'a}nchez and Molina, Jes{\'u}s Garc{\'i}a and Tortosa, Marcos Menarguez},
  booktitle = {Model Driven Architecture -- Foundations and Applications},
  title     = {RubyTL: A Practical, Extensible Transformation Language},
  doi       = {10.1007/11787044_13},
  series    = {ECMDA-FA 2006},
  year      = {2006},
}

@Article{fam2per,
  author   = {Anthony Anjorin and Thomas Buchmann and Bernhard Westfechtel},
  title    = {{{The Families to Persons Case}}},
  series   = {TTC'17},
  year     = {2017},
}

@Standard{QVT2016,
  author   = {OMG},
  revision = {1.3},
  title    = {Meta Object Facility (MOF) 2.0 Query/View/Transformation Specification},
  url      = {https://www.omg.org/spec/QVT/About-QVT/},
  year     = {2016},
}

@InProceedings{Krause2014,
  author    = {Krause, Christian and Tichy, Matthias and Giese, Holger},
  booktitle = {Fundamental Approaches to Software Engineering},
  title     = {Implementing Graph Transformations in the Bulk Synchronous Parallel Model},
  editor    = {Gnesi, Stefania and Rensink, Arend},
  isbn      = {978-3-642-54804-8},
  pages     = {325--339},
  publisher = {Springer Berlin Heidelberg},
  address   = {Berlin, Heidelberg},
  year      = {2014},
}

@Standard{OMG2014,
  author      = {OMG},
  institution = {Object Management Group},
  title       = {{{Object Constraint Language (OCL)}}},
  url         = {https://www.omg.org/spec/OCL/2.4/PDF},
  year        = {2014},
}

@Misc{Hoeppner2022a,
  author    = {Höppner, Stefan and Tichy, Matthias},
  date      = {2022},
  title     = {The Impact of Model Transformation Language Features on Quality Properties of MTLs: A Study Protocol},
  doi       = {10.48550/ARXIV.2209.06570},
  copyright = {Creative Commons Attribution 4.0 International},
  publisher = {arXiv},
}

@Article{Gerpheide2016,
  author       = {Gerpheide, Christine M. and Schiffelers, Ramon R. H. and Serebrenik, Alexander},
  date         = {2016},
  journaltitle = {Software Quality Journal},
  title        = {Assessing and improving quality of QVTo model transformations},
  doi          = {10.1007/s11219-015-9280-8},
  issn         = {1573-1367},
  number       = {3},
  pages        = {797--834},
  url          = {https://doi.org/10.1007/s11219-015-9280-8},
  volume       = {24},
  refid        = {Gerpheide2016},
}

@InProceedings{Kramer2016,
  author    = {Kramer, M. E. and Hinkel, G. and Klare, H. and Langhammer, M. and Burger, E.},
  booktitle = {2nd International Workshop on Human Factors in Modeling, HuFaMo 2016; Saint Malo; France; 4 October 2016 through. Ed. : M. Goulao},
  date      = {2016},
  title     = {A controlled experiment template for evaluating the understandability of model transformation languages},
  language  = {english},
  pages     = {11-18},
  publisher = {{CEUR Workshop Proceedings}},
  series    = {CEUR Workshop Proceedings},
  volume    = {1805},
  issn      = {1613-0073},
}

@InProceedings{Strueber2016,
  author       = {Str{\"u}ber, Daniel and Anjorin, Anthony},
  booktitle    = {HuFaMo@ MoDELS},
  date         = {2016},
  title        = {Comparing Reuse Mechanisms for Model Transformation Languages: Design for an Empirical Study.},
  organization = {Citeseer},
  pages        = {27--32},
}

@InProceedings{Chechik2016,
  author    = {Chechik, Marsha and Famelis, Michalis and Salay, Rick and Str{\"u}ber, Daniel},
  booktitle = {Integrated Formal Methods},
  title     = {Perspectives of Model Transformation Reuse},
  editor    = {{\'A}brah{\'a}m, Erika and Huisman, Marieke},
  isbn      = {978-3-319-33693-0},
  pages     = {28--44},
  publisher = {Springer International Publishing},
  address   = {Cham},
  year      = {2016},
}

@Article{Kusel2015,
  author       = {Kusel, A. and Schönböck, J. and Wimmer, M. and Kappel, G. and Retschitzegger, W. and Schwinger, W.},
  date         = {2015},
  journaltitle = {Software \& Systems Modeling},
  title        = {Reuse in model-to-model transformation languages: are we there yet?},
  doi          = {10.1007/s10270-013-0343-7},
  issn         = {1619-1374},
  number       = {2},
  pages        = {537--572},
  volume       = {14},
  refid        = {Kusel2015},
}
\endgroup

\appendix
\section{USM Results for Moderation Effects}
\label{apdx:results}
See \Cref{tbl:moderation:bx,tbl:moderation:exp,tbl:moderation:explang,tbl:moderation:inc,tbl:moderation:io,tbl:moderation:lang,tbl:moderation:mms,tbl:moderation:mmsanity,tbl:moderation:ms,tbl:moderation:ts}.

\begingroup
\renewcommand{\arraystretch}{1.2}
\begin{table*}
	\caption{Overview of moderation effects of meta-model size}\label{tbl:moderation:mms}
	\begin{tabularx}{\textwidth}{>{\raggedright\arraybackslash\hspace{0pt}}X*{7}{>{\centering\arraybackslash\hspace{0pt}}X}}
		\toprule
		& Comprehensibility& Ease of Writing & Expressiveness  & Tool Support     & Maintainability & Productivity      & Reusability     \\
		\midrule
		Bidirectionality   & 0.0009 & 0.1644 & 0.0004 & 0.0341 & 0.0247 & 0.0197 & 0.0218\\
		Incrementality     & 0.0174 & 0.0003 & 0.0086 & 0.0335 & 0 & 0.0224 & 0.0054\\
		Mappings           & 0.018 & 0.0309 & 0.0058 & 0.0049 & 0 & 0.0012 & 0.0106\\
		Model Management   & 0.1389 & 0.0264 & 0.0006 & 0.023 & 0 & 0.0177 & 0.0199\\
		Model Navigation   & 0.001 & 0.0438 & 0.0019 & 0.0128 & 0.0032 & 0.0422 & 0.018\\
		Model Traversal    & 0.0382 & 0.059 & 0 & 0.0422 & 0.0023 & 0.037 & 0.0106\\
		Pattern Matching   & 0.0091 & 0 & 0.0003 & 0.03 & 0.0028 & 0.0418 & 0.0093\\
		Reuse Mechanisms   & 0.0495 & 0 & 0.0077 & 0.0542 & 0.0693 & 0.0943 & 0.0021\\
		Traceability       & 0.0778 & 0.0464 & 0.00001 & 0.0714 & 0.021 & 0.076 & 0.0223\\
		\bottomrule
	\end{tabularx}
\end{table*}

\begin{table*}
	\caption{Overview of moderation effects of model size}\label{tbl:moderation:ms}
	\begin{tabularx}{\textwidth}{>{\raggedright\arraybackslash\hspace{0pt}}X*{7}{>{\centering\arraybackslash\hspace{0pt}}X}}
		\toprule
		& Comprehensibility& Ease of Writing & Expressiveness  & Tool Support     & Maintainability & Productivity      & Reusability     \\
		\midrule
		Bidirectionality   & 0.1&0.066&0.036&0.048&0.03&0.069&0.012 \\
		Incrementality     & 0.065&0.14&0.019&0.036&0.031&0.052&0.009 \\
		Mappings           & 0.085&0.033&0.076&0.042&0.032&0.083&0.003 \\
		Model Management   & 0.36&0.089&0.059&0.023&0.047&0.079&0.04\\
		Model Navigation   & 0.074&0.007&0.04&0.053&0.05&0.072&0.011\\
		Model Traversal    & 0.125&0.069&0.038&0.048&0.031&0.104&0.008\\
		Pattern Matching   & 0.133&0.052&0.038&0.056&0.024&0.036&0.014\\
		Reuse Mechanisms   & 0.096&0.00009&0.029&0.005&0.11&0.093&0.036\\
		Traceability       & 0.078&0.173&0.02&0.048&0.096&0.108&0.016\\
		\bottomrule
	\end{tabularx}
\end{table*}

\begin{table*}
	\caption{Overview of moderation effects of transformation size}\label{tbl:moderation:ts}
	\begin{tabularx}{\textwidth}{>{\raggedright\arraybackslash\hspace{0pt}}X*{7}{>{\centering\arraybackslash\hspace{0pt}}X}}
		\toprule
		& Comprehensibility& Ease of Writing & Expressiveness  & Tool Support     & Maintainability & Productivity      & Reusability     \\
		\midrule
		Bidirectionality   & 0.1 & 0.15 & 0 & 0.27 & 0.033 & 0.097 & 0.044 \\
		Incrementality     & 0.086 & 0.0075 & 0.0086 & 0.33 & 0.043 & 0.092 & 0.050 \\
		Mappings           & 0.13 & 0.0055 & 0 & 0.17 & 0.042 & 0.16 & 0.034 \\
		Model Management   & 0.32 & 0.094 & 0.021 & 0.23 & 0.012 & 0.13 & 0.056\\
		Model Navigation   & 0.15 & 0.058 & 0.0022 & 0.25 & 0.049 & 0.13 & 0.064\\
		Model Traversal    & 0.069 & 0.045 & 0.0038 & 0.21 & 0.061 & 0.15 & 0.015\\
		Pattern Matching   & 0.16 & 0.055 & 0.040 & 0.28 & 0.075 & 0.12 & 0.055\\
		Reuse Mechanisms   & 0.21 & 0.019 & 0.0077 & 0.16 & 0.27 & 0.23 & 0.087\\
		Traceability       & 0.11 & 0.11 & 0 & 0.25 & 0.067 & 0.12 & 0.040\\
		\bottomrule
	\end{tabularx}
\end{table*}

\begin{table*}
	\caption{Overview of moderation effects of the amount of bidirectional use-cases}\label{tbl:moderation:bx}
	\begin{tabularx}{\textwidth}{>{\raggedright\arraybackslash\hspace{0pt}}X*{7}{>{\centering\arraybackslash\hspace{0pt}}X}}
		\toprule
		& Comprehensibility& Ease of Writing & Expressiveness  & Tool Support     & Maintainability & Productivity      & Reusability     \\
		\midrule
		Bidirectionality   & 0.067 & 0.011 & 0.136 & 0.101 & 0.026 & 0.092 & 0.010 \\
		Incrementality     & 0.036 & 0.005 & 0.120 & 0.115 & 0.003 & 0.099 & 0.019 \\
		Mappings           & 0.029 & 0.034 & 0.065 & 0.084 & 0.002 & 0.072 & 0.001 \\
		Model Management   & 0.143 & 0.048 & 0.175 & 0.066 & 0.010 & 0.053 & 0.014\\
		Model Navigation   & 0.058 & 0.003 & 0.155 & 0.049 & 0.037 & 0.071 & 0.020\\
		Model Traversal    & 0.024 & 0.068 & 0.142 & 0.071 & 0.023 & 0.117 & 0.012\\
		Pattern Matching   & 0.045 & 0.006 & 0.110 & 0.092 & 0.002 & 0.058 & 0.002\\
		Reuse Mechanisms   & 0.040 & 0.026 & 0.130 & 0.220 & 0.100 & 0.190 & 0.013\\
		Traceability       & 0.065 & 0.120 & 0.150 & 0.160 & 0.060 & 0.150 & 0.012\\
		\bottomrule
	\end{tabularx}
\end{table*}

\begin{table*}
	\caption{Overview of moderation effects of developer experience}\label{tbl:moderation:exp}
	\begin{tabularx}{\textwidth}{>{\raggedright\arraybackslash\hspace{0pt}}X*{7}{>{\centering\arraybackslash\hspace{0pt}}X}}
		\toprule
		& Comprehensibility& Ease of Writing & Expressiveness  & Tool Support     & Maintainability & Productivity      & Reusability     \\
		\midrule
		Bidirectionality   & 0.05 & 8.4E-06 & 0.011 & 0.08 & 0.037 & 0.084 & 0.033 \\
		Incrementality     & 0.045 & 0.0003 & 0.007 & 0.083 & 0.032 & 0.055 & 0.046 \\
		Mappings           & 0.00032 & 0 & 0.0013 & 0.072 & 0.052 & 0.1 & 0.039 \\
		Model Management   & 0.15 & 0.025 & 0.032 & 0.085 & 0.05 & 0.073 & 0.034 \\
		Model Navigation   & 0.053 & 0.043 & 0.013 & 0.08 & 0.032 & 0.074 & 0.05 \\
		Model Traversal    & 0.045 & 0.0008 & 0.011 & 0.1 & 0.024 & 0.14 & 0.054 \\
		Pattern Matching   & 0.013 & 0.024 & 0.0052 & 0.081 & 0.032 & 0.078 & 0.048 \\
		Reuse Mechanisms   & 0.047 & 0 & 0.0015 & 0.08 & 0.077 & 0.055 & 0.022 \\
		Traceability       & 0.072 & 0.017 & 0.011 & 0.055 & 0.042 & 0.066 & 0.04 \\
		\bottomrule
	\end{tabularx}
\end{table*}

\begin{table*}
	\caption{Overview of moderation effects of the semantic gap between input and output}\label{tbl:moderation:io}
	\begin{tabularx}{\textwidth}{>{\raggedright\arraybackslash\hspace{0pt}}X*{7}{>{\centering\arraybackslash\hspace{0pt}}X}}
		\toprule
		& Comprehensibility& Ease of Writing & Expressiveness  & Tool Support     & Maintainability & Productivity      & Reusability     \\
		\midrule
		Bidirectionality   & 0.01 & 0.033 & 0.008 & 0 & 0.089 & 0.056 & 0.041 \\
		Incrementality     & 0.006 & 0.067 & 0.009 & 0 & 0.119 & 0.046 & 0.045 \\
		Mappings           & 0.013 & 0.033 & 0.002 & 0 & 0.053 & 0.049 & 0.043 \\
		Model Management   & 0.1 & 0.023 & 0.021 & 0.009 & 0.163 & 0.052 & 0.054 \\
		Model Navigation   & 0.027 & 0.01 & 0.001 & 0.007 & 0.112 & 0.1 & 0.043 \\
		Model Traversal    & 0.023 & 0.032 & 0 & 0.002 & 0.194 & 0.067 & 0.029 \\
		Pattern Matching   & 0.032 & 0.033 & 0 & 0.029 & 0.239 & 0.096 & 0.032 \\
		Reuse Mechanisms   & 0.016 & 0.033 & 0.008 & 0.007 & 0.237 & 0.068 & 0.033 \\
		Traceability       & 0.032 & 0.027 & 0.0003 & 0.012 & 0.184 & 0.058 & 0.038 \\
		\bottomrule
	\end{tabularx}
\end{table*}

\begin{table*}
	\caption{Overview of moderation effects of the sanity of involved meta-models}\label{tbl:moderation:mmsanity}
	\begin{tabularx}{\textwidth}{>{\raggedright\arraybackslash\hspace{0pt}}X*{7}{>{\centering\arraybackslash\hspace{0pt}}X}}
		\toprule
		& Comprehensibility& Ease of Writing & Expressiveness  & Tool Support     & Maintainability & Productivity      & Reusability     \\
		\midrule
		Bidirectionality   & 0.2 & 0.0001 & 0.00002 & 0.035 & 0.025 & 0.074 & 0.045 \\
		Incrementality     & 0.06 & 0.06 & 0.001 & 0.03 & 0.013 & 0.097 & 0.047 \\
		Mappings           & 0.09 & 0.03 & 0.03 & 0.07 & 0.005 & 0.074 & 0.059 \\
		Model Management   & 0.21 & 0.02 & 0.02 & 0.006 & 0.03 & 0.086 & 0.073 \\
		Model Navigation   & 0.11 & 0.12 & 0.002 & 0.06 & 0.04 & 0.099 & 0.044 \\
		Model Traversal    & 0.1 & 0.08 & 0 & 0.08 & 0.02 & 0.13 & 0.069 \\
		Pattern Matching   & 0.09 & 0 & 0.01 & 0.05 & 0.01 & 0.07 & 0.081  \\
		Reuse Mechanisms   & 0.11 & 0 & 0.008 & 0.03 & 0.05 & 0.094 & 0.059 \\
		Traceability       & 0.11 & 0.001 & 0 & 0.05 & 0.05 & 0.14 & 0.079 \\
		\bottomrule
	\end{tabularx}
\end{table*}

\begin{table*}
	\caption{Overview of moderation effects of the amount of incremental use-cases}\label{tbl:moderation:inc}
	\begin{tabularx}{\textwidth}{>{\raggedright\arraybackslash\hspace{0pt}}X*{7}{>{\centering\arraybackslash\hspace{0pt}}X}}
		\toprule
		& Comprehensibility& Ease of Writing & Expressiveness  & Tool Support     & Maintainability & Productivity      & Reusability     \\
		\midrule
		Bidirectionality   & 0 & 0.06 & 0.039 & 0.0045 & 0.034 & 0 & 0.00031 \\
		Incrementality     & 0 & 0.00067 & 0.012 & 0.0042 & 0.079 & 0 & 0.006 \\
		Mappings           & 0 & 0.017 & 0.079 & 0.0018 & 0.061 & 0.0072 & 0.017 \\
		Model Management   & 0.13 & 0.027 & 0.068 & 0.016 & 0.062 & 0.00038 & 0.039 \\
		Model Navigation   & 0 & 0.24 & 0.066 & 0.037 & 0.098 & 0.033 & 0.023 \\
		Model Traversal    & 0 & 0.047 & 0.025 & 0.028 & 0.083 & 0 & 0.0091 \\
		Pattern Matching   & 0.052 & 0 & 0.024 & 0.00048 & 0.073 & 0.032 & 0.034 \\
		Reuse Mechanisms   & 0.047 & 0 & 0.024 & 0.0036 & 0.14 & 0.051 & 0.034 \\
		Traceability       & 0.059 & 0 & 0.088 & 0.005 & 0.13 & 0 & 0.029 \\
		\bottomrule
	\end{tabularx}
\end{table*}

\begin{table*}
	\caption{Overview of moderation effects of the choice of language}\label{tbl:moderation:lang}
	\begin{tabularx}{\textwidth}{>{\raggedright\arraybackslash\hspace{0pt}}X*{7}{>{\centering\arraybackslash\hspace{0pt}}X}}
		\toprule
		& Comprehensibility& Ease of Writing & Expressiveness  & Tool Support     & Maintainability & Productivity      & Reusability     \\
		\midrule
		Bidirectionality   & 0 & 0 & 0.045 & 0 & 0.022 & 0 & 0 \\
		Incrementality     & 0 & 0 & 0.058 & 0 & 0.0035 & 0 & 0 \\
		Mappings           & 0 & 0.0063 & 0.043 & 0 & 0.0047 & 0.0052 & 0 \\
		Model Management   & 0.080 & 0 & 0.0050 & 0.0084 & 0.0077 & 0.0058 & 0 \\
		Model Navigation   & 0 & 0.031 & 0.051 & 0.022 & 0.038 & 0.027 & 0.0015 \\
		Model Traversal    & 0 & 0.00081 & 0.049 & 0.0075 & 0.040 & 0 & 0 \\
		Pattern Matching   & 0.028 & 0.00081 & 0.018 & 3.4E-05 & 0.019 & 0.0076 & 0.0065 \\
		Reuse Mechanisms   & 0.037 & 6.2E-09 & 0.042 & 0.0015 & 0.102 & 0.027 & 0.029 \\
		Traceability       & 0.036 & 0.022 & 0.0042 & 0.026 & 0.024 & 0 & 0.041 \\
		\bottomrule
	\end{tabularx}
\end{table*}

\begin{table*}
	\caption{Overview of moderation effects of the experience in the used languages}\label{tbl:moderation:explang}
	\begin{tabularx}{\textwidth}{>{\raggedright\arraybackslash\hspace{0pt}}X*{7}{>{\centering\arraybackslash\hspace{0pt}}X}}
		\toprule
		& Comprehensibility& Ease of Writing & Expressiveness  & Tool Support     & Maintainability & Productivity      & Reusability     \\
		\midrule
		Bidirectionality   & 0 & 0 & 0 & 0 & 0.0006 & 0 & 0 \\
		Incrementality     & 0 & 0 & 0.0086 & 0 & 0 & 0 & 0 \\
		Mappings           & 0 & 2.4E-17 & 4E-07 & 0 & 0 & 0.0006 & 0 \\
		Model Management   & 0.090 & 0 & 0.021 & 0.024 & 0 & 0.0012 & 0 \\
		Model Navigation   & 0 & 0.008 & 0.0022 & 0.0088 & 0.0014 & 0.043 & 0.015 \\
		Model Traversal    & 0 & 0.0008 & 0 & 0.0065 & 0.0001 & 0 & 0 \\
		Pattern Matching   & 0.0005 & 1.5E-19 & 4.4E-05 & 0.00049 & 0.00069 & 0.0011 & 0.0065 \\
		Reuse Mechanisms   & 0.0056 & 0.0014 & 0.0077 & 0.0071 & 0.018 & 0.025 & 0.0022 \\
		Traceability       & 0.022 & 2E-17 & 4E-06 & 0.028 & 0.014 & 0 & 0.016 \\
		\bottomrule
	\end{tabularx}
\end{table*}
\endgroup
\onecolumn



### How *Hard* or *Easy* is it to *write* model transformations?

❶ Choose one of the following answers
Please choose **only one** of the following:

○ Very Hard

○ Hard

○ Neither Hard nor Easy

○ Easy

○ Very Easy

Ease of writing describes the degree of effectiveness and efficiency with which a developer can produce a transformation for a specific purpose.

### How *Inexpressive* or *Expressive* are the languages used for writing model transformations?

❶ Choose one of the following answers
Please choose **only one** of the following:

○ Very Inexpressive

○ Inexpressive

○ Neither Inexpressive nor Expressive

○ Expressive

○ Very Expressive

Expressiveness describes the degree of effectiveness and efficiency with which language constructs support transformation development.

## How *Unmaintainable* or *Maintainable* are model transformations?

❶ Choose one of the following answers
Please choose **only one** of the following:

◯ Very Unmaintainable

◯ Unmaintainable

◯ Neither Unmaintainable nor Maintainable

◯ Maintainable

◯ Very Maintainable

Maintainability describes the degree of effectiveness and efficiency with which a transformation can be modified.

## How *Unproductive* or *Productive* is developing model transformations?

❶ Choose one of the following answers
Please choose **only one** of the following:

◯ Very Unproductive

◯ Unproductive

◯ Neither Unproductive nor Productive

◯ Productive

◯ Very Productive

Productivity describes the degree of effectiveness and efficiency with which transformations can be developed and used.

### How *Constrained* or *Reusable* are parts of transformations in the languages used for writing model transformations?

❶ Choose one of the following answers
Please choose **only one** of the following:

◯ Very Constrained

◯ Constrained

◯ Neither Constrained nor Reusable

◯ Reusable

◯ Very Reusable

Reuseability describes the degree of effectiveness and efficiency with which parts of a transformation can be reused to create new transformations (with different purpose).

### How B*ad* or G*ood* is the tool support for languages used for writing model transformations?

❶ Choose one of the following answers
Please choose **only one** of the following:

◯ Very Bad

◯ Bad

◯ Neither Bad nor Good

◯ Good

◯ Very Good

Tool Support describes the degree of effectiveness and efficiency with which tools support developers in their effort.

# Capability utilisation of Model Transformation Languages

In the following you will be asked to estimate how often you use certain **capabilities** of model transformation languages.

Each question presents a description of the language capability that is being assessed.

## What percentage of your transformations utilise *Bidirectionality* functionality of the used languages?

❶ Only numbers may be entered in this field.
❶ Your answer must be between 0 and 100
Please write your answer here:

[ ]

Capabilities to define transformations from source to target and target to source in one transformation rule. E.g. relations defined in QvTR can be executed in both directions.

## What percentage of your transformations utilise *Incrementality* functionality of the used languages?

❶ Only numbers may be entered in this field.
❶ Your answer must be between 0 and 100
Please write your answer here:

[ ]

Capabilities to define transformations that are applied whenever the source model changes and only manipulate those parts that did change.

## What percentage of your transformations utilise *Mapping* functionality of the used languages?

❶ Only numbers may be entered in this field.
❶ Your answer must be between 0 and 100
Please write your answer here:

[ ]

Capabilities to explicitly define correspondence between input and output elements. E.g. ATL rules require specification of input type and output type(s).

### What percentage of your transformations utilise *Model Management* functionality of the used languages?

❶ Only numbers may be entered in this field.
❶ Your answer must be between 0 and 100
Please write your answer here:

[ ]

Capabilities to automatically read and write models from and to files or other sources. E.g. for transformation languages built on EMF you can specify the input model and output path without having to handle reading and writing manually.

### What percentage of your transformations utilise *Model Navigation* functionality of the used languages?

❶ Only numbers may be entered in this field.
❶ Your answer must be between 0 and 100
Please write your answer here:

[ ]

Capabilities to seamlessly navigate a given model structure. E.g. OCL provides dedicated model navigation.

### What percentage of your transformations utilise *Model Traversal* functionality of the used languages?

❶ Only numbers may be entered in this field.
❶ Your answer must be between 0 and 100
Please write your answer here:

[ ]

Capabilities to automatically find and apply transformations to model elements. E.g. in ATL you do not have to specify where the elements on which a rule is applied to are taken from. The transformation engine selects those from the input model automatically.

### What percentage of your transformations utilise *Pattern Matching* functionality of the used languages?

❗ Only numbers may be entered in this field.
❗ Your answer must be between 0 and 100
Please write your answer here:

[ ]

Capabilities to automatically match patterns of model elements and apply transformations to them. E.g. Henshin allows you to define a graph structure that is matched in the source model during transformation.

### What percentage of your transformations utilise *Reuse* functionality of the used languages?

❗ Only numbers may be entered in this field.
❗ Your answer must be between 0 and 100
Please write your answer here:

[ ]

Capabilities to enable reusing whole transformations or parts of transformations. E.g. any type of rule inheritance. Copying parts of your code does not constitute reuse.

### What percentage of your transformations utilise *Tracing* functionality of the used languages?

❗ Only numbers may be entered in this field.
❗ Your answer must be between 0 and 100
Please write your answer here:

[ ]

Capabilities to (automatically) generate and maintain trace links between source and target elements. E.g. ATL automatically resolves traces to the corresponding output model element when an input model element is referenced as the value for an attribute.

# Experience

How many *years* have you worked on developing model transformations?

❗ Only numbers may be entered in this field.

Please write your answer here:

How many *hours* do you work on developing model transformations *per month*?

❗ Only numbers may be entered in this field.

Please write your answer here:

### Which *5* of the following languages do you *use most often* to develop model transformations?

❶ Check all that apply
❶ Please select at most 5 answers
Please choose **all** that apply:

- [ ] AGG
- [ ] ATL
- [ ] eMoflon
- [ ] ETL
- [ ] Fujaba
- [ ] GReAT
- [ ] GrGen
- [ ] Henshin
- [ ] Java
- [ ] JavaScript
- [ ] JTL
- [ ] Kermeta
- [ ] QVTo
- [ ] QVTr
- [ ] Ruby
- [ ] RubyTL
- [ ] Tefkat
- [ ] Viatra
- [ ] Xtend
- [ ] Other 1
- [ ] Other 2
- [ ] Other 3
- [ ] Other 4
- [ ] Other 5

### Name the *first* other language you are using for developing model transformations.

Only answer this question if the following conditions are met:
((LANGS_Other1.NAOK (/limesurvey/index.php/questionAdministration/view/surveyid/112652/gid/121/qid/1722) == "Y"))

Please write your answer here:

### Name the *second* other language you are using for developing model transformations.

Only answer this question if the following conditions are met:
((LANGS_Other2.NAOK (/limesurvey/index.php/questionAdministration/view/surveyid/112652/gid/121/qid/1722) == "Y"))

Please write your answer here:

### Name the *third* other language you are using for developing model transformations.

Only answer this question if the following conditions are met:
((LANGS_Other3.NAOK (/limesurvey/index.php/questionAdministration/view/surveyid/112652/gid/121/qid/1722) == "Y"))

Please write your answer here:

## Name the *fourth* other language you are using for developing model transformations.

Only answer this question if the following conditions are met:
((LANGS_Other4.NAOK (/limesurvey/index.php/questionAdministration/view/surveyid/112652/gid/121/qid/1722) == "Y"))

Please write your answer here:

## Name the *fifth* other language you are using for developing model transformations.

Only answer this question if the following conditions are met:
((LANGS_Other5.NAOK (/limesurvey/index.php/questionAdministration/view/surveyid/112652/gid/121/qid/1722) == "Y"))

Please write your answer here:

# For how many *years* have you *used* these *languages*?

❶ Only numbers may be entered in these fields.
❶ Each answer must be at least 0

Please write your answer(s) here:

AGG

ATL

eMoflon

ETL

Fujaba

GReAT

GrGen

Henshin

Java

JavaScript

JTL

Kermeta

QVTo

QVTr

Ruby

RubyTL

Tefkat

Viatra

Xtend

Other1

Other2

Other3

Other4

Other5

# For how many *hours* do you *use* these *languages per month*?

❶ Only numbers may be entered in these fields.
❶ Each answer must be at least 0
Please write your answer(s) here:

AGG

ATL

eMoflon

ETL

Fujaba

GReAT

GrGen

Henshin

Java

JavaScript

JTL

Kermeta

QVTo

QVTr

Ruby

RubyTL

Tefkat

Viatra

Xtend

Other1

Other2

Other3

Other4

Other5

How *many elements* do the meta-models involved in your transformations have? Please estimate the percentage of your use cases that fall in the following ranges.

❶ Each answer must be between 0 and 100
❶ The sum must be at most 100
❶ Only integer values may be entered in these fields.
Please write your answer(s) here:

#elements ≤ 10

☐

10 < #elements ≤ 20

☐

20 < #elements ≤ 50

☐

50 < #elements ≤ 100

☐

100 < #elements ≤ 1.000

☐

#elements > 1.000

☐

E.g. If half of the meta-models in your transformations have 25 elements and the other half are meta-models with 4 elements, you would put 50 for #elements ≤ 10 and 50 for 20 < #elements ≤ 50.

How *large* are the models you transform measured in *number of model elements*? Please estimate the percentage of your use cases that fall in the following ranges.

❶ Each answer must be between 0 and 100
❶ The sum must be at most 100
❶ Only integer values may be entered in these fields.
Please write your answer(s) here:

#elements ≤ 10

[ ]

10 < #elements ≤ 100

[ ]

100 < #elements ≤ 1.000

[ ]

1.000 < #elements ≤ 10.000

[ ]

10.000 < #elements ≤ 100.000

[ ]

#elements > 100.000

[ ]

E.g. if 1/3 of all models you tranform contain 200 elements and the rest are larger than 100.000 elements, you would put 33 for 100 < #elements ≤ 1.000 and 66 for #elements > 100.000.

How *Small* or *Large* are your transfomrations? Please estimate the percentage of your use cases that fall in the following ranges.

❶ Each answer must be between 0 and 100
❶ The sum must be at most 100
❶ Only integer values may be entered in these fields.
Please write your answer(s) here:

Tiny (e.g. LOC ≤ 100)

☐

Small (e.g.100 < LOC ≤ 500)

☐

Medium (e.g. 500 < LOC ≤ 1.000)

☐

Large (e.g. 1.000 < LOC ≤ 5.000)

☐

Very Large (e.g. 5.000 < LOC ≤ 10.000)

☐

Huge (e.g. LOC : 10.000)

☐

How *Dissimilar* or *Similar* in structure are input and output meta-models in your transformations? Please estimate the percentage of your use cases that fall in the following ranges.

❶ Each answer must be between 0 and 100
❶ The sum must be at most 100
❶ Only integer values may be entered in these fields.
Please write your answer(s) here:

very dissimilar

[　　　]

dissimilar

[　　　]

neither similar nor dissimilar

[　　　]

similar

[　　　]

very similar

[　　　]

The structure of a meta-model is define by the number of elements and their associations with each other.

How *Dissimilar* or *Similar* are the attribute types of input and output elements that are related to each other in your transformations? Please estimate the percentage of your use cases that fall in the following ranges.

❶ Each answer must be between 0 and 100
❶ The sum must be at most 100
❶ Only integer values may be entered in these fields.
Please write your answer(s) here:

very dissimilar

[ ]

dissimilar

[ ]

neither similar nor dissimilar

[ ]

similar

[ ]

very similar

[ ]

E.g. when mapping a Class to a Table and assigning the ClassName as the TableName the attribute types are identical when both are Strings.

How *Bad* or *Well structured* are the meta-models in your transformations? Please estimate the percentage of your use cases that fall in the following ranges.

❶ Each answer must be between 0 and 100
❶ The sum must be at most 100
❶ Only integer values may be entered in these fields.
Please write your answer(s) here:

very bad

bad

neither well nor bad

well

very well

A well structured meta-model does for example not split related data over a large number of meta-model elements if it can be avoided.

How *Bad* or *Well documented* are the meta-models in your transformations? Please estimate the percentage of your use cases that fall in the following ranges. Only consider documentation of the meta-model itself not documentation in you code.

❶ Each answer must be between 0 and 100
❶ The sum must be at most 100
❶ Only integer values may be entered in these fields.
Please write your answer(s) here:

very bad

[          ]

bad

[          ]

neither well nor bad

[          ]

well

[          ]

very well

[          ]

Documentation means description of the meta-model elements, their attributes and associations as well as any invariants on them.

---

What percentage of your use cases require *Synchronization* between "input" and "output".

❶ Only numbers may be entered in this field.
❶ Your answer must be between 0 and 100
Please write your answer here:

[          ]

> ### What percentage of your use cases require *Incrementality*?
>
> ❗ Only numbers may be entered in this field.
> ❗ Your answer must be between 0 and 100
> Please write your answer here:
>
> [              ]

Thank you very much for participating in this survey! We appreciate you took the time.

# Good bye and have a nice day!

02.02.2023 – 12:56

Submit your survey.
Thank you for completing this survey.


\twocolumn
\section{Mail Templates}
\label{apdx:mails}

Dear \$\{Author Name\},

We found your contact information while conducting a structured literature search on model-to-model transformations. We seek your expertise on that topic.

We invite you to participate in our study `The Impact of Model Transformation Language Capabilities on the Perception of Language Quality Properties` (see below for the URL). It will only take about 20-25 minutes to complete our online survey. We believe your answers can provide meaningful insights and help drive the field further.

Our survey is based on a large-scale interview study that qualitatively assessed what the community believes to be the main factors that drive the advantages and disadvantages of Model Transformation Languages for model-to-model transformations.
Our results have been published in the Empirical Software Engineering journal \url{https://doi.org/10.1007/s10664-022-10194-7}

Our survey now quantifies our results to provide a clear picture of which of our identified factors are most important. The methodology for this survey has been peer reviewed at the Registered Reports track at ESEM’22 and is available under \url{https://doi.org/10.48550/arXiv.2209.06570}

The survey is available at

\url{https://sp2.informatik.uni-ulm.de/limesurvey/index.php/112652?lang=en}

It will be open till January 15, 2023.

All responses are completely anonymous.

Many thanks for supporting our research!

Best regards

Stefan Höppner
\section{Data Privacy Agreement}
\label{apdx:consent_form}

To invite people to participate in this survey, we used author information (first name, last name and email address) from published academic papers in the domain of model driven software engineering. We will delete your personal information 2 months after you received the invitation.

The participation in this online survey is anonymous. Your name and email address are only used to invite you.

For future publications, we will publish and further process the anonymous raw data collected in this survey. This includes aggregating and statistical analysis of answers provided by participants. We will perform this in a way that does not allow inferring the identity of individual participants (e.g., by stripping free text answers of information that could identify a participant).
Contact Information

If you have any questions about this survey or the data you provided, please contact us:

\textit{Stefan Höppner}\\
\textit{stefan.hoeppner@uni-ulm.de}\\
\textit{Institute of Software Engineering and Programming Languages,}\\
\textit{Ulm University,}\\
\textit{James-Franck-Ring, 89069 Ulm, Germany}


\end{document}